\documentclass[modern]{aastex631}
\usepackage[T1]{fontenc}

\newcommand{\ldl}{$\lambda/\Delta\lambda$}
\newcommand{\teff}{$T_{\rm eff}$}
\newcommand{\logg}{$\log{g}$}

\newcommand{\vtan}{$v_{\rm tan}$}
\newcommand{\sname}{J1249+3621}
\newcommand{\msun}{M$_{\odot}$}

\received{May 28, 2024}
\revised{\today}
\accepted{July 8, 2024}

\submitjournal{ApJL}

\shorttitle{A Hypervelocity L Subdwarf}
\shortauthors{Burgasser et al.}

\graphicspath{{./}{}}

\begin{document}

\title{Discovery of a Hypervelocity L Subdwarf at the Star/Brown Dwarf Mass Limit}

\correspondingauthor{Adam Burgasser}
\email{aburgasser@ucsd.edu}

\author[0000-0002-6523-9536]{Adam J.\ Burgasser}
\affiliation{Department of Astronomy \& Astrophysics, UC San Diego, La Jolla, CA, USA}

\author[0000-0003-0398-639X]{Roman Gerasimov}
\affiliation{Department of Physics \& Astronomy, University of Notre Dame, Notre Dame, IN 46556, USA}

\author[0000-0002-4086-3180]{Kyle Kremer}
\affiliation{TAPIR, California Institute of Technology, Pasadena, CA 91125, USA}

\author[0000-0002-5253-0383]{Hunter Brooks}
\affiliation{Department of Astronomy and Planetary Science, Northern Arizona University, Flagstaff, AZ 86011, USA}

\author[0009-0005-8159-8490]{Efrain Alvarado III}
\affiliation{Department of Astronomy, University of California, Berkeley, CA 94720-3411, USA}

\author[0000-0002-6294-5937]{Adam C.\ Schneider}
\affiliation{US Naval Observatory, Flagstaff Station, Flagstaff, AZ, USA}

\author[0000-0002-1125-7384]{Aaron M.\ Meisner}
\affiliation{NSF National Optical-Infrared Astronomy Research Laboratory, 950 North Cherry Avenue, Tucson, AZ 85719, USA}

\author[0000-0002-9807-5435]{Christopher A.\ Theissen}
\affiliation{Department of Astronomy \& Astrophysics, UC San Diego, La Jolla, CA, USA}

\author[0000-0002-1420-1837]{Emma Softich}
\affiliation{Department of Astronomy \& Astrophysics, UC San Diego, La Jolla, CA, USA}

\author[0000-0002-1480-9041]{Preethi Karpoor}
\affiliation{Department of Astronomy \& Astrophysics, UC San Diego, La Jolla, CA, USA}

\author[0000-0003-2235-761X]{Thomas P. Bickle}
\affil {School of Physical Sciences, The Open University, Milton Keynes, MK7 6AA, UK}
\affiliation{Backyard Worlds: Planet 9}

\author[0000-0003-4905-1370]{Martin Kabatnik}
\affiliation{Backyard Worlds: Planet 9}

\author[0000-0003-4083-9962]{Austin Rothermich}
\affiliation{Department of Astrophysics, American Museum of Natural History, Central Park West at 79th Street, NY 10024, USA}
\affiliation{Department of Physics, Graduate Center, City University of New York, 365 5th Ave., New York, NY 10016, USA}
\affiliation{Department of Physics and Astronomy, Hunter College, City University of New York, 695 Park Avenue, New York, NY, 10065, USA}

\author[0000-0001-7896-5791]{Dan Caselden}
\affiliation{Department of Astrophysics, American Museum of Natural History, Central Park West at 79th Street, NY 10024, USA}

\author[0000-0003-4269-260X]{J.\ Davy Kirkpatrick}
\affiliation{IPAC, Mail Code 100-22, Caltech, 1200 East California Boulevard, Pasadena, CA 91125, USA}

\author[0000-0001-6251-0573]{Jacqueline K.\ Faherty}
\affiliation{Department of Astrophysics, American Museum of Natural History, Central Park West at 79th Street, NY 10024, USA}

\author[0000-0003-2478-0120]{Sarah L.\ Casewell}
\affiliation{Centre for Exoplanet Research, School of Physics and Astronomy
University of Leicester, University Road, Leicester, LE1 7RH, UK}

\author[0000-0002-2387-5489]{Marc J.\ Kuchner}
\affiliation{Exoplanets and Stellar Astrophysics Laboratory, NASA Goddard Space Flight Center, 8800 Greenbelt Road, Greenbelt, MD 20771, USA}

\collaboration{100}{The Backyard Worlds: Planet 9 Collaboration}

\begin{abstract}
We report the discovery of a high velocity, very low-mass star or brown dwarf whose kinematics suggest it is unbound to the Milky Way. CWISE~J124909.08+362116.0 was identified by citizen scientists {in} the Backyard Worlds: Planet 9 program as a high proper motion ($\mu$ $=$ 0$\farcs$9/yr) faint red source. Moderate resolution spectroscopy with Keck/NIRES reveals it to be a metal-poor early L subdwarf with a large radial velocity ($-$103$\pm$10~km/s), and its estimated distance of 125$\pm$8~pc yields a 
speed of 456$\pm$27~km/s in the Galactic rest frame, near the local escape velocity for the Milky Way. 
{We explore several potential scenarios for the origin of this source, including 
ejection from the Galactic center 
$\gtrsim$3~Gyr in the past, 
survival as the mass donor companion to an exploded white dwarf.
acceleration through a three-body interaction with a black hole binary in a globular cluster, and accretion from a Milky Way satellite system.} 
CWISE~J1249+3621 is the first hypervelocity very low mass star or brown dwarf to be found, and the nearest of all such systems.
{It} may represent a broader population of very high velocity, low-mass objects that have undergone extreme accelerations.
\end{abstract}


\section{Introduction} \label{sec:intro}

The majority of stars in the neighborhood of the Sun have low relative velocities ($v$ $\approx$ 10-30~km/s) reflecting their common origin in star forming clusters concentrated in the plane of the Milky Way. 
A rare subset of nearby stars have much higher velocities 
({\vtan} $\gtrsim$ 400~km/s; $<$0.3\% of stars {within 1 kpc}; \citealt{2015ApJ...813...26F}). 
{These stars may originate from the Milky Way's ancient halo population,
or underwent strong dynamical interactions with compact objects 
such as the Milky Way's central supermassive black hole \citep{1988Natur.331..687H}
of compact binaries in dense clusters \citep{2003ApJ...599.1129Y,2019MNRAS.489.4543F},
or may be the survivors of the supernova explosion of a binary companion \citep{1961BAN....15..265B,2000ApJ...544..437P}.
The fastest ``hypervelocity'' stars are unbound to the Milky Way's gravitational potential and 
may even have extragalactic origins \citep{2009ApJ...691L..63A,2011A&A...535A..70P}.}
These rare objects trace extreme interactions that may be explored through their trajectories, velocity distributions, and atmospheric properties \citep{2015ARA&A..53...15B}.

The Gaia mission \citep{2021A&A...650C...3G} has greatly expanded our sample of high-velocity stars by providing 5D (position, {parallax}, and proper motion) or 6D (plus radial velocity or RV) coordinates for billions of stars out to kiloparsec {distances}. These measurements, combined with detailed chemical abundances from 
RAVE \citep{2006AJ....132.1645S},
LAMOST \citep{2012RAA....12.1197C},
APOGEE \citep{2017AJ....154...94M}, 
and other spectral surveys have enabled the discovery and characterization
of over a dozen hypervelocity stars (e.g., \citealt{2019ApJS..244....4D,2022AJ....164..187Q,2023ApJ...944L..39L,2024arXiv240210714S}) {that} originate from environments as diverse as the Galactic center, globular clusters, or satellite systems. 
Current studies focus on deep optical measurements of rare and distant stars, and primarily sample main sequence and red giant stars over a limited range of mass (0.7~M$_\odot$ $\lesssim$ M $\lesssim$ 2~M$_\odot$) and age ($\lesssim$10~Gyr for high-velocity OBAFG stars) which may limit our ability to probe compositions and origins.

The citizen science project {\em Backyard Worlds: Planet 9} (BYW; \citealt{2017ApJ...841L..19K}) takes advantage of multi-epoch infrared photometry and astrometry from the {\em Wide-field Infrared Survey Explorer} ({\em WISE}; \citealt{2010AJ....140.1868W}) and its extended {\em NEOWISE} mission \citep{2014ApJ...792...30M} to search for
faint, infrared moving sources identified by a community of citizen scientists.
BYW is ideally designed to find low-mass, high velocity stars and brown dwarfs,
including local low-temperature metal-poor subdwarfs---the L, T, and Y subdwarfs---from the thick disk and halo populations
\citep{2020ApJ...898...77S,2020ApJ...899..123M,2021ApJ...915..120M,2021ApJ...915L...6K,2022AJ....163...47B,burgasser2024}.
In this Letter, we report the discovery of a nearby, metal-poor L subdwarf, CWISE~J124909.08+362116.0 (hereafter {\sname}) whose speed may exceed the local escape velocity of the Milky Way, making it the first low-mass hypervelocity star 
{and the nearest such system to the Sun.}

\section{Identification and Spectral Observations} \label{sec:discobs}

{\sname} was identified by citizen scientists Tom Bickle, Martin Kabatnik, and Austin Rothermich in multi-epoch unWISE images \citep{2014AJ....147..108L,2018AJ....156...69M,2019ApJS..240...30S} 
on the BYW citizen science portal.\footnote{{\url{http://www.backyardworlds.org}.}} Its $W2$ magnitude and $J-W2$ color (Table~\ref{tab:source}) 
suggest an early-type L dwarf at an estimated distance of $\approx$100~pc \citep{2021ApJS..253....7K}.
Combining astrometry from PanSTARRS \citep{2016arXiv161205560C} and the UKIDSS Hemisphere Survey (UHS; \citealt{2018MNRAS.473.5113D})
yields a proper motion of $\mu$ = 884$\pm$5~mas yr$^{-1}$,
suggesting a tangential velocity of {\vtan} $\approx$ 420~km s$^{-1}$ and making it a high priority target for spectroscopic followup.

\begin{deluxetable}{lcl}
\tablecaption{Properties of J1249+3621 \label{tab:source}} 
\tabletypesize{\scriptsize} 
\tablehead{ 
\colhead{Property} & 
\colhead{Value} & 
\colhead{Reference}  
}
\startdata
$\alpha_{J2000}$  & 12$^h$49$^m$09$\fs$08   & 1 \\
$\delta_{J2000}$   & +36$\degr$21$\arcmin$16$\arcsec$   & 1  \\
$\mu_\alpha\cos\delta$   & 344$\pm$5~mas/yr   & 2,3 \\
$\mu_\delta$   & $-$814$\pm$5~mas/yr   & 2,3  \\
$i$ (AB) & 21.48$\pm$0.15~mag & 2 \\	
$z$ (AB) & 20.01$\pm$0.06~mag & 2 \\	
$y$ (AB) & 19.13$\pm$0.05~mag & 2 \\	
$J$ (Vega) & 17.10$\pm$0.03~mag & 3 \\
$K$ (Vega) & 16.46$\pm$0.04~mag & 3 \\
$W1$ (Vega) & 15.92$\pm$0.04~mag & 1 \\
$W2$ (Vega) & 15.59$\pm$0.07~mag & 1 \\
SpT & sdL1 & 4 \\
$d_{est}$\tablenotemark{a} & 125$\pm$8~pc & 4 \\
{\vtan} & 524$\pm$33~km/s & 4 \\
{\teff}\tablenotemark{b} & 1715~K to 2320~K & 4,5,6 \\
{\logg}\tablenotemark{b} & 4.4 to 5.1 (cm/s$^2$) & 4,5,6 \\
{[M/H]}\tablenotemark{b} & $-$1.4 to $-$0.5 & 4,5,6 \\
{[$\alpha$/Fe]} & +0.25$\pm$0.07 & 4,5 \\
Est.\ Mass &  0.082$^{+0.002}_{-0.003}$~M$_\odot$ & 4,7 \\
$RV$ & $-$103$\pm$10~km/s & 4 \\
$U_{LSR}$\tablenotemark{c} & 449$\pm$28~km/s & 4 \\
$V_{LSR}$\tablenotemark{c} & $-$292$\pm$19~km/s & 4 \\
$W_{LSR}$\tablenotemark{c} & $-$15$\pm$11~km/s & 4 \\
$v_{GRF}$\tablenotemark{d} & 456$\pm$27~km/s & 4 \\
\enddata
\tablenotetext{a}{ {Estimated from the spectral classification, $JKW1W2$ photometry, and the spectral type/absolute magnitude relations of \citet{2018ApJ...864..100G} and \citet{2019MNRAS.486.1260Z}}.}
\tablenotetext{b}{ Based on the $\pm$1$\sigma$ range of Elf Owl and SAND model fits.}
\tablenotetext{c}{ Local Standard of Rest (LSR) velocities {assuming a solar motion from} \citet{2010MNRAS.403.1829S}.}
\tablenotetext{d}{ Galactic rest frame (GRF) speed assuming $v_{circ}$ = 220~km/s at the Solar radius.}
\tablerefs{
(1) CatWISE2020 \citep{2021ApJS..253....8M} {at astrometric epoch 2015 May 28 (UT)};
(2) PanSTARRS \citep{2016arXiv161205560C};
(3) UKIRT Hemisphere Survey \citep{2018MNRAS.473.5113D};
(4) This paper;
(5) \citet{Alvarado_2024};
(6) \citet{2024ApJ...963...73M};
(7) \citet{2024arXiv240501634G}
}
\end{deluxetable}


{\sname} was observed on 30 January 2024 (UT) in clear and windy conditions with the Near-Infrared Echellette Spectrometer (NIRES; \citealt{2004SPIE.5492.1295W}) on the Keck II 10m telescope, a cross-dispersed spectrograph that provides {\ldl} $\approx$ 2700 spectra over 
0.9--2.45 $\mu$m.
We obtained six exposures of 300~s each at an average airmass of 1.06 with the slit aligned with the parallactic angle. Exposures were made in an ABBA pattern, nodding 10$\arcsec$ along the slit for background subtraction. We also observed the A0~V star HD~108140 ($V$ = 9.35) at a similar airmass, and dome flat lamp exposures at the start of the night for pixel response calibration. Data were reduced using a modified version of the Spextool package
\citep{2004PASP..116..362C}, following the procedure of \citet{2003PASP..115..389V} for flux calibration and telluric absorption correction.

\section{Analysis} \label{sec:analysis}

\subsection{Classification and Atmosphere Parameters} \label{sec:classification}

Figure~\ref{fig:spectrum} compares a smoothed version of our spectrum to near-infrared spectral standards from the SpeX Prism Library Analysis Toolkit (SPLAT; \citealt{2017ASInC..14....7B}). {\sname} exhibits the characteristic features of L-type dwarfs, with strong H$_2$O absorption at 1.4~$\mu$m and 1.9~$\mu$m, and FeH, Na~I, and K~I absorption in the 1.0--1.3~$\mu$m region. Its NIR spectral slope is distinctly bluer than normal L dwarf spectra, and its 2.3~$\mu$m CO band is highly suppressed indicating enhanced H$_2$ collision-induced absorption (CIA) in a low-metallicity, low-temperature atmosphere \citep{1969ApJ...156..989L,2003ApJ...592.1186B}. 
Indeed, the spectrum of {\sname} {best matches} that of the L subdwarf 2MASS~J17561080+2815238 (sdL1; \citealt{2010ApJS..190..100K}) but is somewhat bluer, while not as blue as the extreme L subdwarf WISE~J043535.80+211509.2 (esdL1; \citealt{2014ApJ...787..126L,2017MNRAS.464.3040Z}). We classify {\sname} as sdL1 based on comparison to a broad range of dwarf and subdwarf spectra.
This classification, WISE $W1W2$ and UHS $JK$ {Vega} magnitudes, and the spectral type/absolute magnitude relations for L sudwarfs from \citet{2018ApJ...864..100G} and \citet{2019MNRAS.486.1260Z} allow us to estimate a spectrophotometric distance\footnote{This estimate takes into account photometric uncertainties, uncertainties in the absolute magnitude/spectral type relations, and a $\pm$1 subtype uncertainty on the classification. We used the uncertainty-weighted mean across all bands and both relations, which are in formal agreement within the uncertainties.} of 125$\pm$8~pc for {\sname}, {implying} {\vtan} = 524$\pm$33~km/s.

To further evaluate its physical properties, we compared the smoothed spectrum of {\sname} to the Sonora Elf Owl \citep{2024ApJ...963...73M} and Spectral ANalog of Dwarfs (SAND; \citealt{Alvarado_2024}) atmosphere models. These models encompass the temperatures (1500~K $\lesssim$ {\teff} $\lesssim$ 2400~K) and subsolar metallicities ([M/H] $\lesssim$ $-$0.5) of L subdwarfs, and contain up-to-date opacities and treatments for condensation and disequilibrium chemistry.
We used a Metropolis-Hastings Markov Chain Monte Carlo (MCMC) fitting algorithm \citep{1953JChPh..21.1087M,HASTINGS01041970} to fit the models to the apparent spectral flux densities of {\sname}, following the procedure described in \citet{burgasser2024}. Our best fit models and parameters are {shown} in Figure~\ref{fig:spectrum}.
The Elf Owl and SAND grids yield {distinctly different} effective temperatures ({\teff} = 2260$\pm$60~K versus 1785$\pm$70~K) and metallicities ([M/H] = $-$0.63$\pm$0.10 versus $-$1.28$\pm$0.10), but similar surface gravities ({logg} = 4.66$\pm$0.11 versus 4.88$\pm$0.24). 
We note that the Elf Owl grid does not extend to [M/H] $<$ $-$1 and does not include a prescription for condensate cloud formation, although the latter may be less important in metal-poor L subdwarf atmospheres \citep{2007ApJ...657..494B,2021ApJ...923...19G}. 
The best-fit Elf Owl model is a marginally better fit to the observed spectrum,\footnote{The best fits yield $\chi^2_r$ = 9.4 for Elf Owl and $\chi^2_r$ = 14.4 for SAND, where $\chi^2_r$ = $\chi^2$/DOF, with DOF = 263. These values indicate a relative probability $\ln\mathcal{P}$ = $-\chi^2_{r,SAND}/\chi^2_{r,ElfOwl}$ = $-$1.5 or $\mathcal{P}$ = 0.22 that the models equally represent the data.} and its scaled surface fluxes are in good agreement with the spectrophotometric distance estimate ($d$ = 102~pc) 
assuming a radius of 0.08~R$_\odot$.
The model fit discrepancies could be resolved by a direct distance measurement.
{Nevertheless, this analysis confirms} our interpretation of {\sname} as a low-temperature, metal-poor object. 
We note that the SAND models suggest significant alpha enrichment ([$\alpha$/Fe] = +0.25$\pm$0.07)
and the Elf Owl models marginal C/O enrichment (C/O = 0.71$\pm$0.16), {potential} clues to the origin of this source. 

{Adopting generous parameter ranges} of 1715~K $\lesssim$ {\teff} $\lesssim$ 2320~K and
$-$1.4 $\leq$ [M/H] $\leq$ $-$0.5, and {assuming} an age $\tau$ $\geq$ 5~Gyr, the SAND evolutionary models (SANDee; \citealt{2024arXiv240501634G}) predict a mass of 0.082$^{+0.002}_{-0.003}$~M$_\odot$, placing this source {marginally} above the metallicity-dependent Hydrogen Burning Minimum Mass (HBMM $\approx$ 0.080~M$_\odot$ for [M/H] = $-$1). 
The relatively narrow uncertainty range indicated by the evolutionary models is due to the steep decline in temperatures below the HBMM for old low-temperature sources, and does not account for potential systematic biases (e.g., non-solar abundance patterns). 
We conclude that {\sname} is likely a low-mass, {metal-poor} star, with a 10\% probability of being a high-mass brown dwarf.

\newbox{\bigpicturebox}
\begin{figure}
\centering
\sbox{\bigpicturebox}{%
  \includegraphics[width=.48\textwidth]{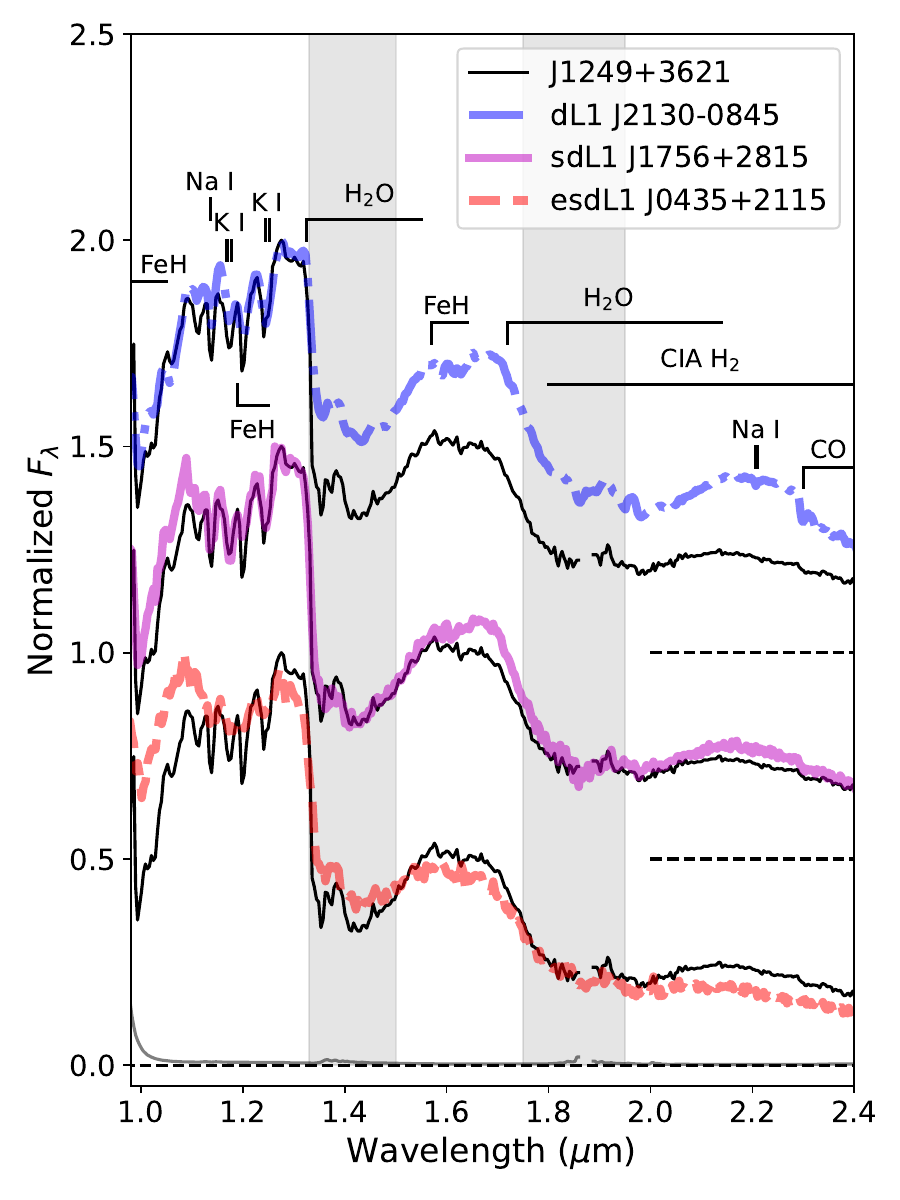}%
}
\usebox{\bigpicturebox}\hfill
\begin{minipage}[b][\ht\bigpicturebox][s]{.48\textwidth}
\includegraphics[width=\textwidth]{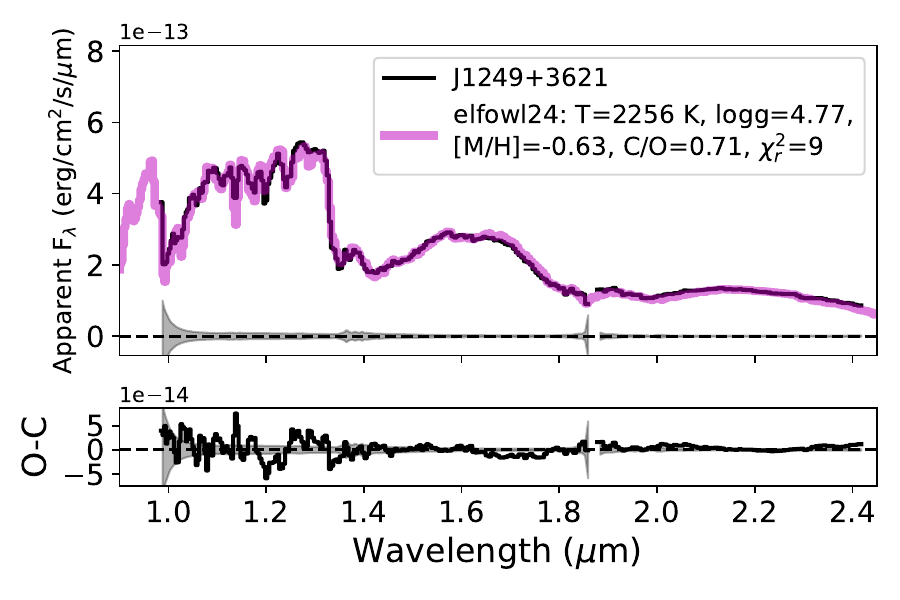}\vfill
\includegraphics[width=\textwidth]{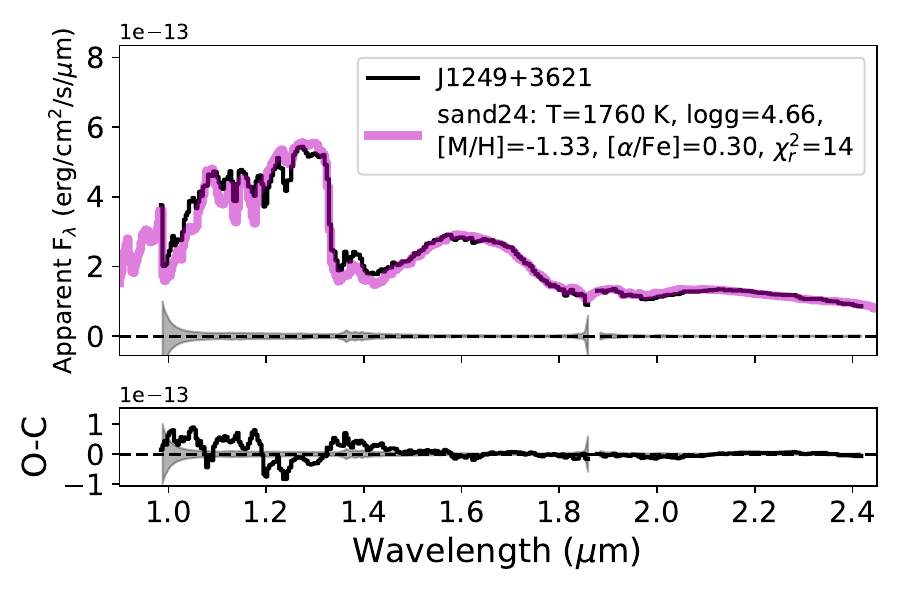}
\end{minipage}
\caption{
(Left): Keck/NIRES spectrum of {\sname} smoothed to a resolution of {\ldl} = 150 (black lines), compared to the L dwarf templates
2MASSW J2130446-084520 (L1, blue dot-dashed line; data from \citealt{2014ApJ...794..143B}),
2MASS~J17561080+2815238 (sdL1, magenta line; data from \citealt{2010ApJS..190..100K}),
and WISE~J043535.80+211509.2 (esdL1, red dashed line; data from \citealt{2014ApJ...787..126L}).
The spectrum of {\sname} is normalized at 1.3~$\mu$m and the comparison spectra normalized to maximize agreement in the 1.0--1.3~$\mu$m range. The dwarf and subdwarf comparisons are offset for ease of comparison.
Major spectral features are labeled, as are regions of strong telluric absorption at 1.35--1.5~$\mu$m and 1.75--1.85~$\mu$m (vertical grey bands).
(Right)
Comparison of the smoothed spectrum of {\sname} (black lines) to best fit models (magenta lines) from the
Sonora Elf Owl (top; \citealt{2024ApJ...963...73M}) 
and SAND (bottom; \citealt{Alvarado_2024}) atmosphere grids. 
Spectra are scaled to apparent fluxes using the UHS $J$-band magnitude of {\sname}, and the models scaled to minimize $\chi^2_r$.
Model parameters are listed in the figure captions.
Difference spectra (observed minus computed; black line) are compared to the $\pm$1$\sigma$ uncertainty of the observed flux densities (grey band) in the bottom panels.
\label{fig:spectrum}}
\end{figure}

\subsection{Radial Velocity and Kinematics} \label{sec:kinematics}

\begin{figure}
\centering
\includegraphics[width=0.48\textwidth]{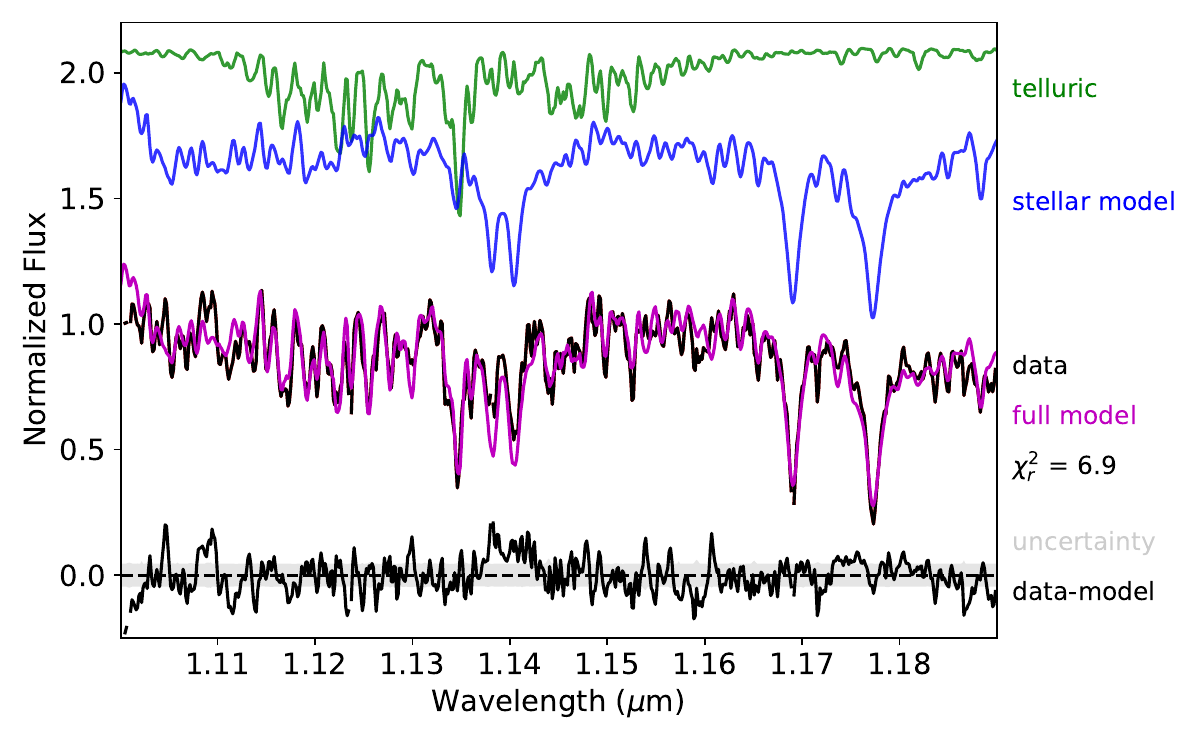}
\includegraphics[width=0.48\textwidth]{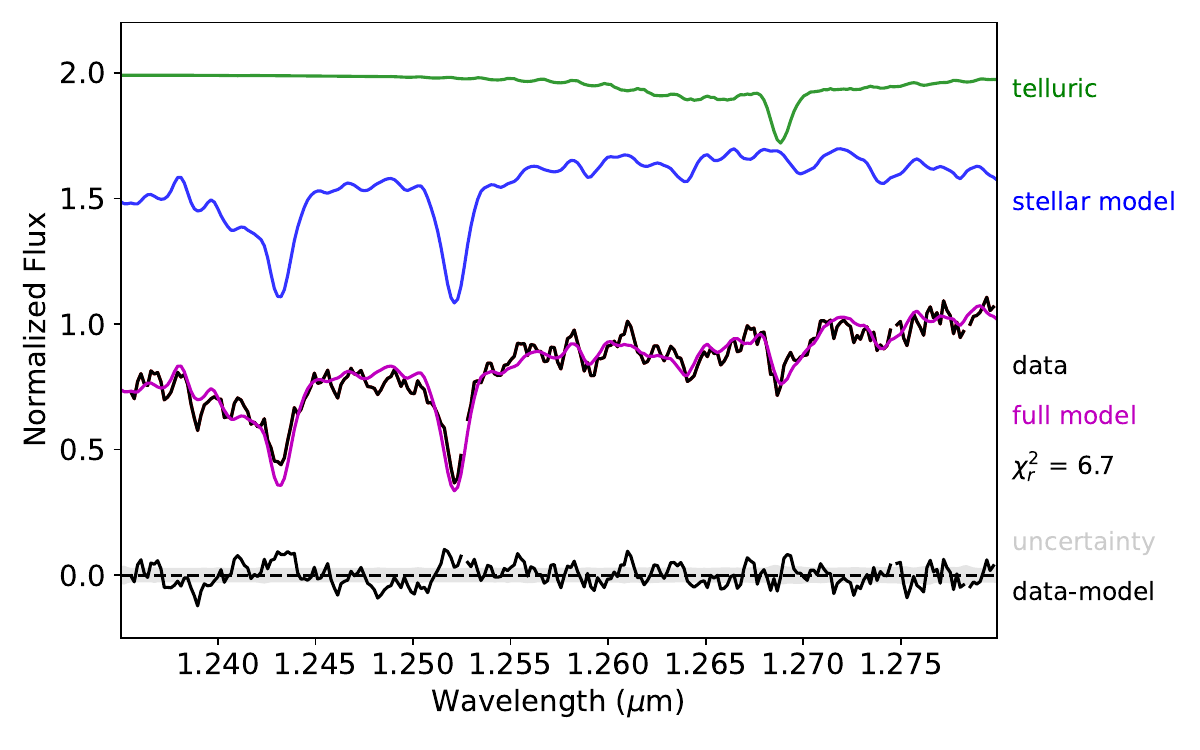}
\caption{
Forward modeling of the Keck/NIRES spectrum of {\sname} in the 
1.10--1.19~$\mu$m (left), and
1.235--1.28~$\mu$m (right) spectral regions, both of  which contain stellar and telluric absorption features.
Each panel displays from top to bottom: 
the telluric spectrum in green, the stellar model in blue,
the combined model in magenta overlaid on the observed spectrum in black,
and the difference spectrum (data-model) in black overlaid on the $\pm$1$\sigma$ uncertainty band in grey.
The reduced $\chi^2$ of each fit is indicated in the text to the right of the plot.
\label{fig:rv}}
\end{figure}

At full resolution, the Keck/NIRES data permit assessment of the radial velocity of {\sname},
which we approached using a forward modeling technique described in \citet{burgasser2024}.
In brief, we fit the extracted NIRES spectrum without telluric correction to a high resolution stellar atmosphere model, $M(\lambda)$, from \citet[BT-Settl]{2012RSPTA.370.2765A} and
an empirical telluric absorption template, $T(\lambda)$, from \citet{1991aass.book.....L} using the parameterized data model
\begin{equation}
    D(\lambda+\delta_\lambda) = C(\lambda)\times\left[\left(M(\lambda^*)\otimes\kappa_R(v\sin{i})\right){\times}T^\alpha(\lambda)\right]\otimes\kappa_G(v_b)+\delta_f.
\end{equation}
Here, $C(\lambda)$ is a fifth-order polynomial continuum correction, 
$\kappa_R(v\sin{i})$ is a rotational broadening profile for projected velocity $v\sin{i}$,
$\alpha$ is a scaling exponent for the strength of the telluric absorption,
$\kappa_G({v_b})$ is a Gaussian instrumental broadening profile parameterized by velocity width $v_b$,
and $\delta_\lambda$ and $\delta_f$ represent small offsets to the wavelength scale and normalized flux density to account for residual calibration errors. 
We used a {\teff} = 2000~K, {\logg} = 5.0, solar metallicity model evaluated at a shifted wavelength $\lambda^* = \lambda\left(1+\frac{RV+V_{bary}}{c}\right)$, accounting for the unknown RV and known barycentric motion $v_{bary}$ = 16.82~km s$^{-1}$ at the time of observation.
After identifying an optimized set of parameters using the Nelder-Mead algorithm
\citep{nelder65} with a fixed $v\sin{i}$ = 50~km s$^{-1}$, we used a Metropolis-Hastings MCMC algorithm to map the parameter uncertainty space for the five remaining free parameters: RV, $\alpha$, $v_b$, $\delta_\lambda$ and $\delta_f$.
Fits were conducted in two wavelength regions that contain both stellar and telluric absorption features:
1.10--1.19~$\mu$m which contains Na~I (1.138, 1.140~$\mu$m) and K~I (1.169, 1.177~$\mu$m) stellar lines and a telluric complex over 1.11--1.15~$\mu$m,
and 
1.235--1.28~$\mu$m which contains K~I stellar lines (1.243, 1.252~$\mu$m) and a telluric feature at 1.269~$\mu$m.
We avoided the CO band at 2.3~$\mu$m, commonly used for RV forward modeling of L dwarfs (cf.\ \citealt{2010ApJ...723..684B,2010ApJ...711.1087K,2015AJ....149..104B}), {due to the suppression of this feature by H$_2$ absorption}.
Figure~\ref{fig:rv} displays the best fit models for these regions, which yield 
consistent values for RV ($-$92$^{+13}_{-14}$~km/s and $-$114$^{+13}_{-14}$~km/s).
We adopt a mean RV = $-$103$\pm$10~km/s for {\sname}.

Combining the measured RV, proper motion, and position of {\sname} with its estimated distance, we computed $UVW$ velocities in the Local Standard of Rest (LSR).\footnote{LSR velocity components assume a right-handed coordinate system centered on the Sun with $U$ pointed radially inward, $V$ pointed in the direction of Galactic rotation, and $W$ pointed toward the north Galactic pole. {We assumed} solar velocity components of ($U_{\odot}$, $V_{\odot}$, $W_{\odot}$) = (11.1~km/s, 12.24~km/s, 7.25~km/s) \citep{2010MNRAS.403.1829S}.} 
The velocities (Table~\ref{tab:source}) indicate a slightly retrograde motion relative to Galactic rotation
($V_{LSR}$ = $-$292$\pm$19~km/s), with a trajectory directed radially inward 
($U_{LSR}$ = 449$\pm$28~km/s) and constrained to the Galactic disk
($W_{LSR}$ = $-$15$\pm$11~km s$^{-1}$). 
Assuming a local Galactic circular velocity of $v_{circ}$ = 220~km/s, the velocity of {\sname} translates into a Galactic rest frame speed of $v_{GRF}$ = 456$\pm$27~km/s, or 0.47$\pm$0.03 kpc/Myr.
The median speed is just below the Galactic escape velocity at the Solar radius, with current estimates ranging from
521$^{+46}_{-30}$~km/s \citep[1.6$\sigma$ above]{2017MNRAS.468.2359W}
to 580$\pm$63~km/s \citep[1.8$\sigma$ above]{2018A&A...616L...9M}.
Given the uncertainties in the inferred velocities and potential models, 
we find that {\sname} has a significant probability of being unbound to the Milky Way. 
%
%


\begin{figure}
\centering
\includegraphics[width=0.45\textwidth]{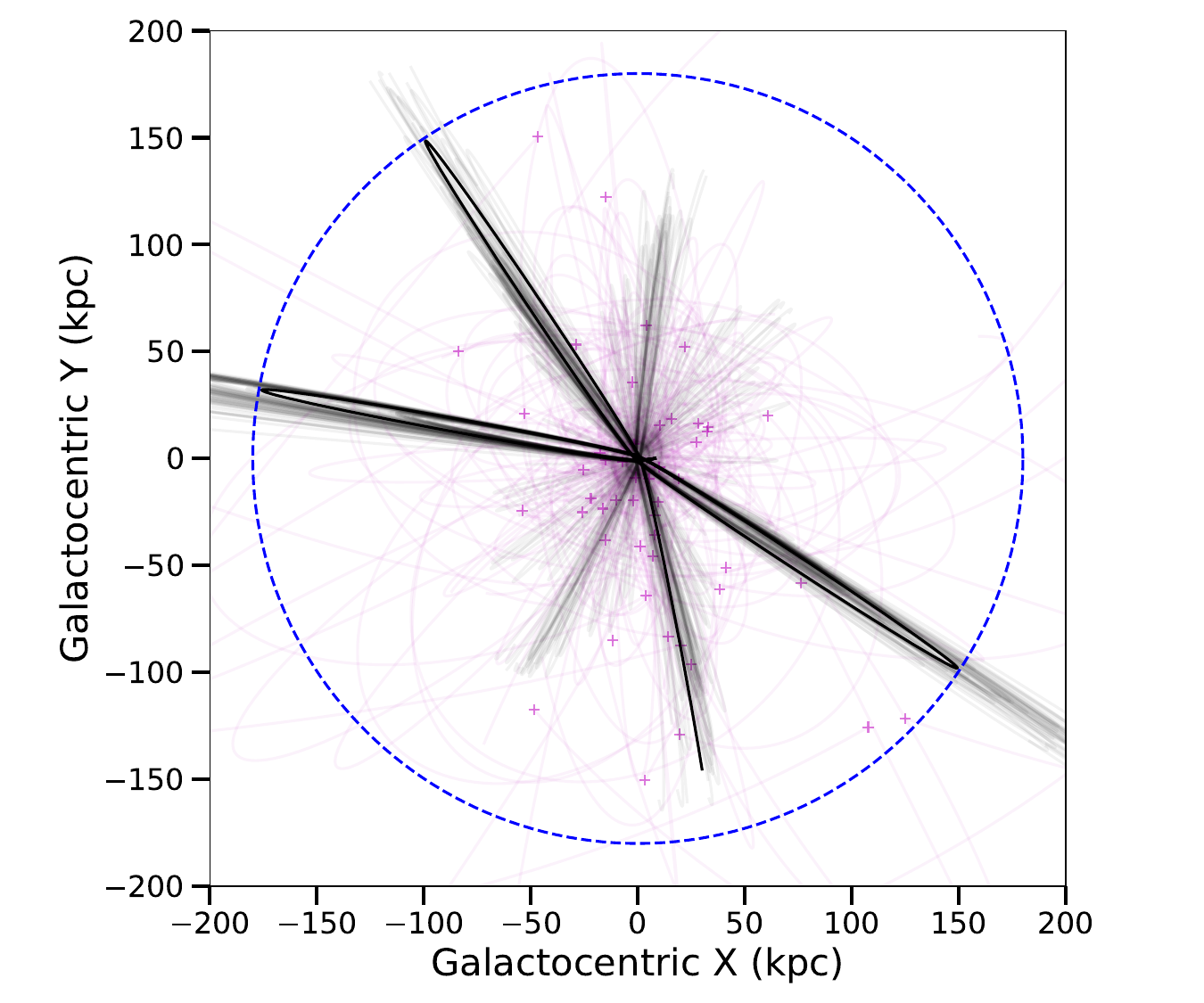}
\includegraphics[width=0.45\textwidth]{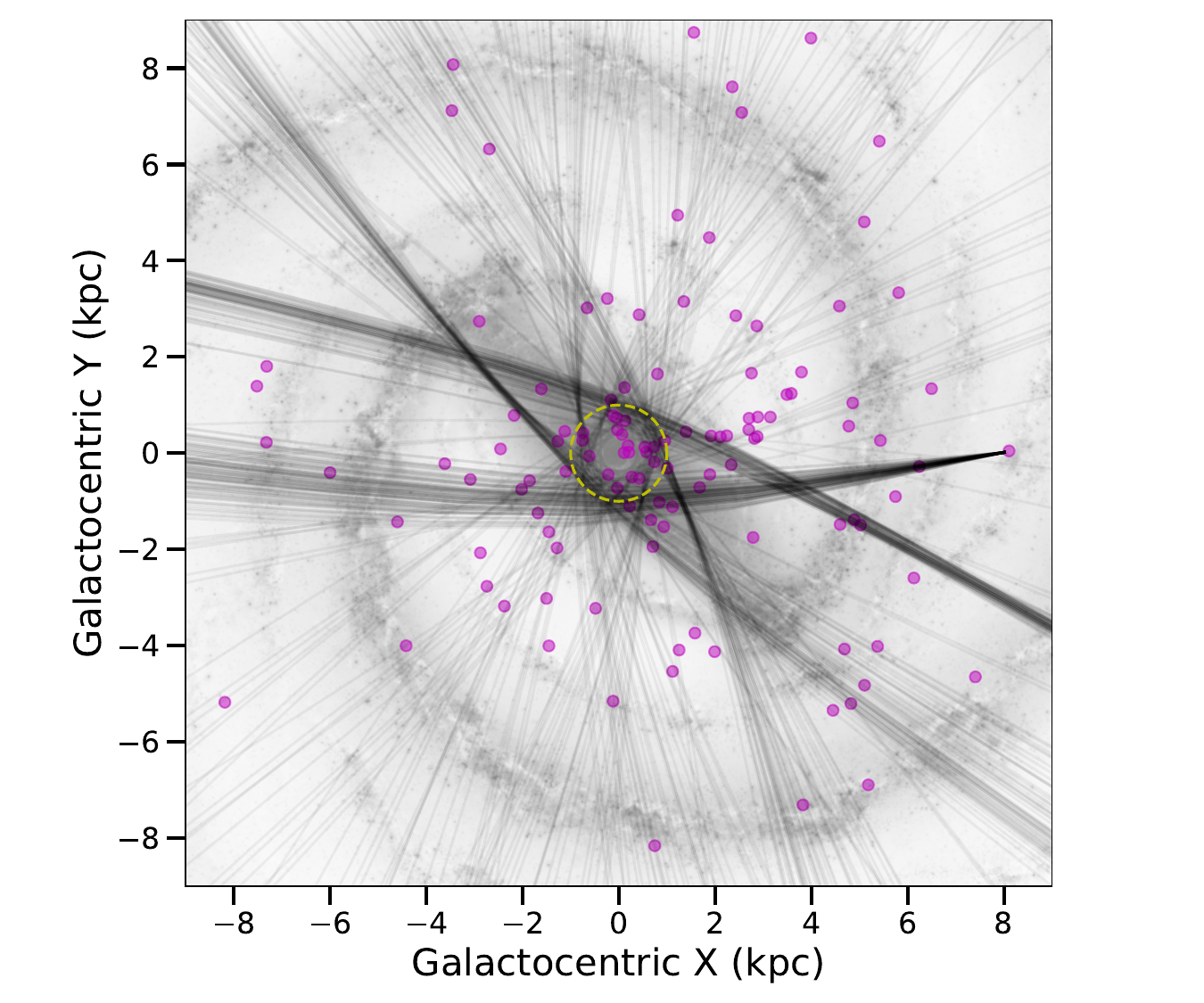} \\
\includegraphics[width=0.45\textwidth]{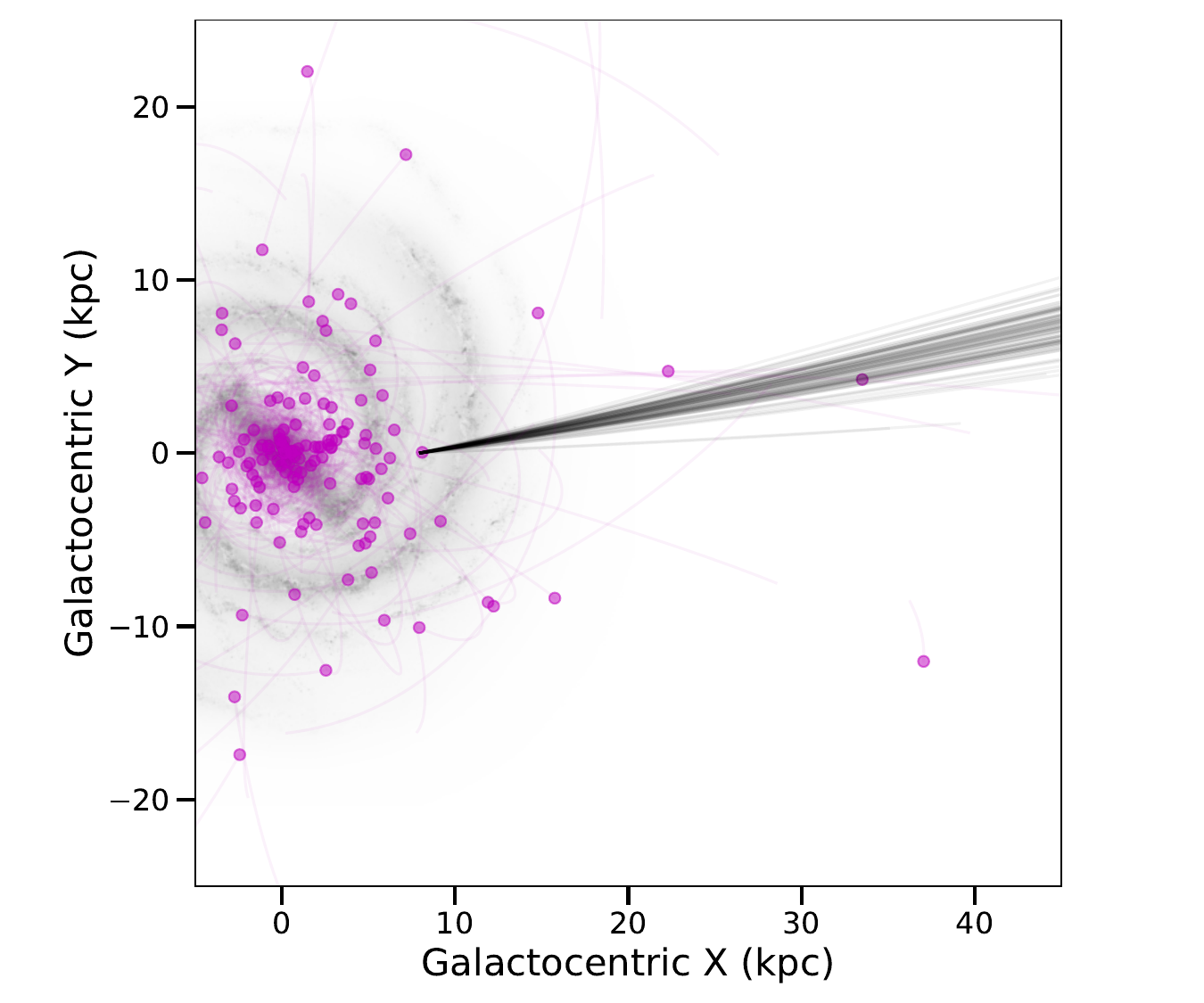}
\includegraphics[width=0.45\textwidth]{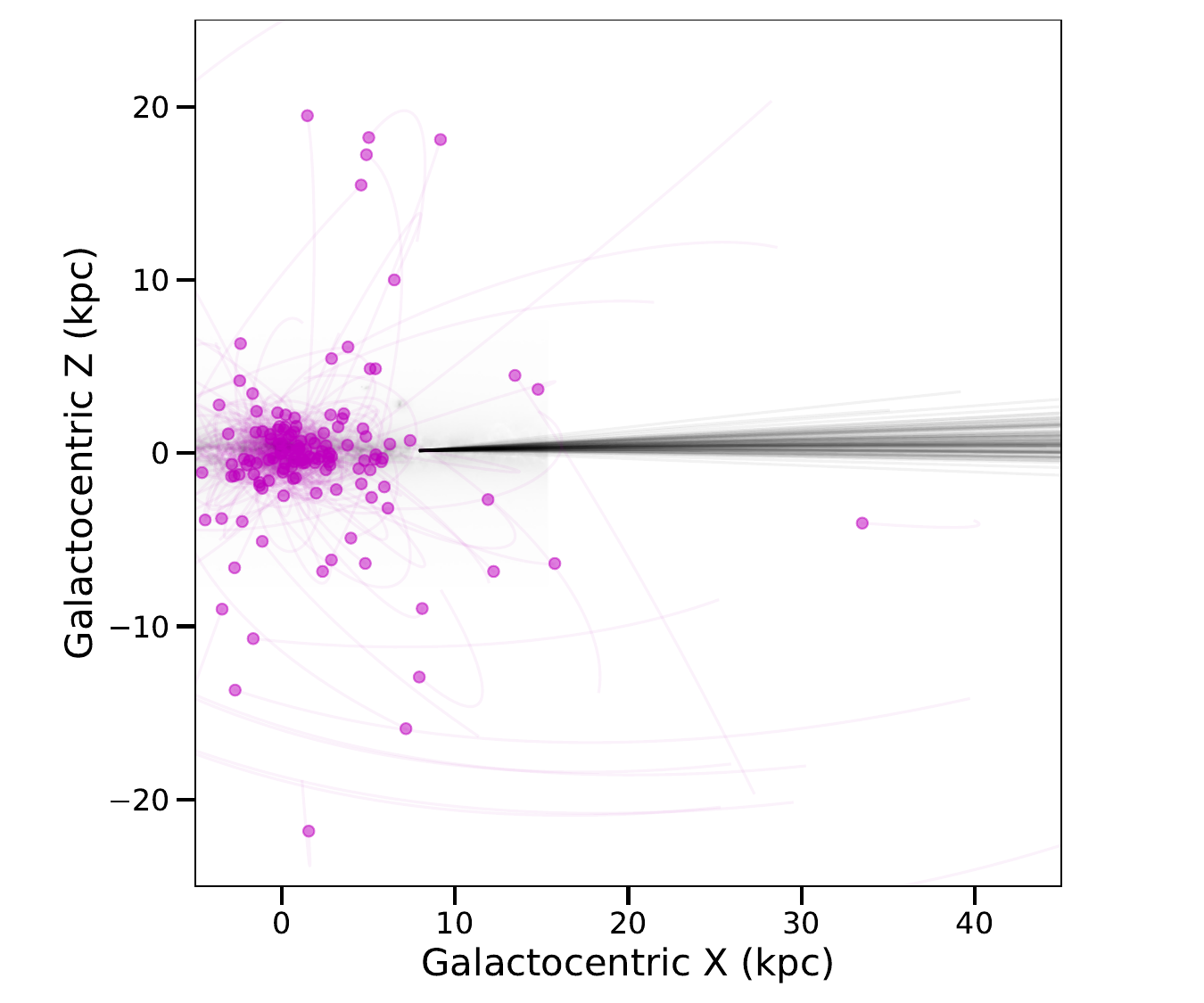} \\
\caption{
Projected orbit of {\sname} from {\tt galpy} \citep{2015ApJS..216...29B} {for 100 initial conditions} sampling uncertainties in distance, proper motion, and RV (black lines).
The top panels display the forward orbit over {10}~Gyr projected onto the Galactic plane on wide (left) and narrow (right) distance scales. 
{The bottom panels display the backward orbit over 150~Myr projected onto the Galactic plane (left) and XZ coordinates (right).} 
The top left panel shows the present-day positions (magenta crosses) and forward-projected orbits (magenta lines) of 50 satellite galaxies from \citet{2018A&A...619A.103F}, 
{as well as the Milky Way's virial radius of 180~kpc (dashed blue circle; \citealt{2023ApJ...945....3S}). 
The top right panel shows the present-day positions of 161 globular clusters from \citet[][magenta circles]{2019MNRAS.484.2832V} and the closest-approach radius of $\sim$1~kpc from the Galactic center (yellow dashed circle).
Globular clusters and their backward-projected orbits (magenta lines) are also shown in the bottom panels.}
\label{fig:orbit}}
\end{figure}

The Galactic orbit of {\sname} was generated using {\tt galpy} \citep{2015ApJS..216...29B}. We used the {axisymmetric} MWPotential2014 potential to integrate the trajectory of {\sname} forward and backward in time by up to {10}~Gyr, with a finer sampling of the backward orbit up to 150~Myr. We drew {100} random initial conditions sampling the uncertainties on distance, proper motion, and radial velocity assuming independent Gaussian distributions. Figure~\ref{fig:orbit} displays the forward and backward trajectories projected onto the disk plane and in cylindrical coordinates.
The forward motion of {\sname} shows a close approach to the inner region of the Milky Way, coming within 0.94$^{+0.28}_{-0.19}$~kpc of the Galactic center, then extending beyond the Milky Way's virial radius of 180~kpc \citep{2023ApJ...945....3S}. 
The median model remains bound to the Milky Way on a $\sim${3}~Gyr, highly eccentric orbit, but 17\% of our simulated orbits are unbound over 10~Gyr. 
The backward orbit is approximately radial and tightly confined to the Galactic plane,
converging to within 2$\degr$ of 05$^h$22$^m$25$^s$ +38$\degr$37$\arcmin$00$\arcsec$ (galactic coordinate 38$\fd$6 +1$\fd$2) by 50~Myr in the past.



\section{Assessing the Origins of J1249+3621} \label{sec:origins}

\subsection{Was J1249+3621 Ejected from the Galactic Center?} \label{sec:cen}

{
{\sname} has a unique trajectory and speed; less than 0.002\% of stars in Gaia within 200~pc of the Sun have comparable tangential velocities.\footnote{This statistic is based on a search of Gaia DR3 for sources with $\pi$ $>$ 5~mas, $\pi/\sigma_\pi$ $>$ 10, and {\vtan} $>$ 500~km/s, which comprises 34 sources out of 2,234,316 without a tangential velocity constraint.} 
While this could nevertheless represent the extreme tail of the halo velocity distribution \citet{2018MNRAS.481.1028H}, 
we explored potential origins of {\sname} in the context of currently known hypervelocity stars.
Its small but nonzero orbital angular momentum in the Galactic rest frame ($L_z$ = 572$^{+147}_{-146}$~kpc~km~s$^{-1}$) and inward trajectory would seem to argue against ejection from the Galactic center through the traditional Hill's mechanism \citep{1988Natur.331..687H}.
However, if {\sname} is bound to the Milky Way, which is the case for 83\% of our orbit simulations, we could be observing it on a return pass after intervals of roughly 3~Gyr,
with torques imparted by asymmetries in the Galactic potential; i.e., spiral structure or the inner bar \citep{2002MNRAS.336..785S,2018MNRAS.476.1561D}.
Moreover, while the majority of Galactic center stars are metal-rich
\citep{2000ApJ...530..307C,2000ApJ...537..205R,2007ApJ...669.1011C}, recent studies have identified metal-poor, alpha-enhanced M giants in the nuclear region similar in nature to {\sname} \citep{2015A&A...584A..45S,2020A&A...642A..81S}.
A small number of metal-poor hypervelocity stars have also been associated with ejection from the Galactic center
\citep{2012ApJ...744L..24L,2023AJ....166...12L}.
Thus, despite its radially inward trajectory, ejection via the Hill's mechanism is a possible origin for {\sname}.
}


\subsection{Is J1249+3621 the Surviving Companion of a Type Ia Supernova?} \label{sec:wd}

{An alternative disk origin for {\sname}} is as the tight binary companion to an accreting white dwarf that 
exceeded the 1.4~M$_\odot$ Chandrasekhar limit \citep{1931MNRAS..91..456C,1935MNRAS..95..207C}, underwent a thermonuclear explosion, and released {\sname} at high speed
{\citep{1961BAN....15..265B,2000ApJ...544..437P,2018ApJ...865...15S}.} 
Several {accreting short-period white dwarf-brown dwarf pairs (polars and cataclysmic variables) are known with} white dwarf masses extending up to 0.94~M$_\odot$ (\citealt{2019MNRAS.484.2566L} and references therein). If {\sname} was the {initially more massive} donor in such a system, its subsequent ejection speed would exceed 690$\left(\frac{P}{1~hr}\right)^{-1/3}$~km s$^{-1}$ based on orbital motion alone, where $P$ is the orbit period at detonation. Periods of 1--1.5~hr are sufficient for Roche lobe overflow\footnote{{Following \citet{1983ApJ...268..368E}, $a$/R$_*$ $\approx$ 3--10 for a star-white dwarf system that evolves from 0.6~{\msun}+0.9~{\msun} to 0.1~{\msun}+1.4~{\msun}, where $a$ is the orbit semimajor axis and R$_*$ the radius of the mass donor. In the latter configuration, P $\approx$ 1--1.5~hr for average densities of 50--80~g~cm$^{-3}$.}} for donors near the HBMM, based on theory and as observed for low-mass cataclysmic variables (e.g., \citealt{1999MNRAS.309.1034K,2006Sci...314.1578L}).
This scaling law yields ejection speeds of 550--700~km s$^{-1}$, on par with {\sname}'s LSR speed of 534~km/s. Contributions from supernova shockwaves and mass stripping could drive ejection velocities 
even higher \citep{2012ApJ...750..151P,2022ApJ...933...38R}.

{As the ejection direction in this scenario is isotropic, the probability of an individual source such as {\sname} passing by the Sun is very low. However, the overall higher density of stars in the Galactic plane makes it more likely that we would see a closely-passing ejectee with a trajectory confined to the plane. We note that}
there are no known supernova remnants in the projected past position of {\sname}, but
as remnants dissipate and merge with the interstellar medium within $\lesssim$1~Myr \citep{2017AJ....153..239L} this a relatively weak constraint on the time since this source could have been ejected.

\subsection{Was J1249+3621 Ejected from a Globular Cluster?} \label{sec:gc}

Another possible origin {for {\sname}} is dynamical ejection from a globular cluster (GC). 
The top-heavy present-day mass functions of these clusters is evidence of efficient tidal dispersion of low-mass objects \citep{2002ApJ...570..171F}, and
large velocity kicks exceeding the 10--100~km/s escape velocities of GCs are a natural outcome of strong three- and four-body dynamical interactions that are commonplace in these systems \citep{1991AJ....101..562L}. Such kicks {are} amplified by interactions with black hole binaries \citep{2023ApJ...953...19C}, which are now understood to dominate the dynamics of the centers of most {GCs} \citep{2020IAUS..351..357K}. The characteristic kick velocity imparted to a star of mass $m$ during a strong encounter with a binary with components of equal mass $M$ and semi-major axis $a$ is roughly:
\begin{equation}
    \label{eq:v_kick}
    v_{\rm kick} \approx \sqrt{ \frac{GM^2}{m a}} \approx 600 \Bigg(\frac{M}{20\,M_{\odot}} \Bigg) \Bigg(\frac{m}{0.1\,M_{\odot}} \Bigg)^{-1/2} \Bigg(\frac{a}{10\,\rm{au}} \Bigg)^{-1/2}\,\rm{km/s}.
\end{equation}
Thus, for $a\lesssim 10\,$au and/or compact object masses $M\gtrsim20\,M_{\odot}$, sufficiently high velocities can be achieved.

To test this mechanism, we searched for {low-mass} high velocity stars in the \texttt{CMC Cluster Catalog} \citep{2020ApJS..247...48K}, a public suite of $N$-body simulations intended to serve as a proxy 
for the Galactic GC population \citep{1996AJ....112.1487H}. These simulations adopt a \citet{2001MNRAS.322..231K} initial mass function ranging from $0.08-150\,M_{\odot}$, and therefore include stars of mass comparable to {\sname}.
At times $t>8\,$Gyr, typical of the ages of Milky Way GCs, we identify roughly $4,000$ low-mass stars (M $<$ 0.2~M$_{\odot}$) ejected with velocities of at least $100\,$km/s, corresponding to a rate of roughly 1~Myr$^{-1}$ across the Milky Way. Of these, six stars are ejected with $V > 500\,$km/s, corresponding to a rate of roughly 2~Gyr$^{-1}$. all six hypervelocity subdwarfs awere  ejected via dynamical encounters with stellar-mass black hole binaries with properties similar to the scales in  Eqn.~\ref{eq:v_kick}.
Hence, simulations support the scenario of GC ejection as a means of generating very high velocity, very low mass objects, albeit as exceedingly rare events.
{The isotropic distribution of ejections further reduces the probability of this scenario, with the same caveat that the higher concentration of GCs in the Galactic plane makes detection of closely-passing ejectees on planar orbits more likely.}


There are no known globular clusters within 5$\degr$ of the projected past position of {\sname}. However, given the long timescales involved it is necessary to account for cluster motion. We used our {\tt galpy} orbits to assess whether the trajectory of {\sname} intersected with any of the 161 GCs in the kinematic catalog of \citet{2019MNRAS.484.2832V}, projecting all orbits back 150~Myr.
{The closest approaches by NGC~3201 and Palomar 1 are more than 4~kpc in separation, making these improbable origin sites.}
The low Galactic latitude of the projected past position of {\sname} ($b$ = +1$\fd$2)
{could argue for an origin from an as yet undiscovered GC hidden in the Galactic plane. Alternately, the higher concentration of GCs near the Galactic center implies that {\sname} could have been ejected from one of these systems, and is now making a return pass $\gtrsim$3~Gyr later. Despite these considerations, we find the GC ejection scenario less likely than the prior two scenarios.}

For completeness, we note that four open clusters lie within 1$\degr$ of the projected past position of {\sname},
including the well-studied NGC~1857 system at 3.1~kpc \citep{1864RSPT..154....1H}. However, the lower stellar densities of these systems and lack of compact binaries from massive progenitors makes dynamic ejection unlikely, and none of these clusters have significantly subsolar metallicities characteristic of {\sname}.

\subsection{Is J1249+3621 an Extragalactic Star?} \label{sec:extragal}

As the bound orbits of {\sname} extend beyond the Milky Way's virial radius, it is possible that this source could have an extragalactic origin, specifically accretion from one of the Milky Way satellites \citep{2009ApJ...691L..63A,2011A&A...535A..70P}. Its present trajectory does not align with distant extragalactic systems such as M31 or the Magellanic Clouds. We examined the intersection of {\sname}'s orbit with 50 Milky Way satellites from \citet{2018A&A...619A.103F} using the same procedure as our GCs, integrating back to {10}~Gyr. Only one system, Tucana III \citep{2015ApJ...813..109D}, comes within 5~kpc of the median orbit of {\sname} about 6~Gyr ago. The substantial uncertainties of the trajectories of {\sname} and the known dwarf satellites over Gyr timescales
means we cannot strictly rule out this scenario; however, we find it {the least likely of the scenarios considered here, given} the trajectory of {\sname} being {confined} to the Galactic plane. 

{Inferring the true origin of {\sname} will require further investigation into its physical and atmospheric properties.
A Galactic center origin requires closer examination of its orbital trajectory
through refinement of its distance (by direct parallax measurement) and velocity components, as well as a more realistic, non-axisymmetric Galactic potential model.
A more detailed compositional analysis would also help clarify its origin.
For example, } 
if {\sname} is the surviving companion of a Type Ia supernova, {its atmosphere may be} enriched with heavy elements, particularly nickel, {depending on the degree of} mass stripping by the supernova blast wave 
\citep{2022ApJ...933...38R}. 
Similarly, if {\sname} was ejected from {the Galactic center,} a globular cluster, or a satellite system, its detailed abundances may provide the {chemical} fingerprint of its origin. 
Better assessment of composition of through additional optical and infrared spectra, and improved atmosphere models exploring specific abundances (e.g., \citealt{2022ApJ...930...24G}) are needed to infer a chemical-based origin.

Finally, we note that at least one other metal-poor L subdwarf, the esdL1 ULAS~J231949.36+044559.5 \citep{2018MNRAS.480.5447Z}, has a high enough estimated tangential velocity (513$^{+50}_{-46}$~km/s) to make it a promising hypervelocity candidate, although no RV has yet been reported for this source.
The existence of at least one and possibly two L subdwarfs within $\sim$200~pc of the Sun with hypervelocity speeds suggests a considerably larger population of such sources could exist in the Milky Way system.

\begin{acknowledgments}
We thank Keck Observatory staff Randy Campbell and Max Service for their support during the Keck/NIRES observations. 
{Figure 3 was made with the MWplot package created by Henry Leung (\url{https://github.com/henrysky/milkyway_plot}) and graphics created by Robert Hurt (\url{https://www.spitzer.caltech.edu/image/ssc2008-10a-a-roadmap-to-the-milky-way}).
The authors acknowledge helpful discussions with Boris Gaensickle and Keith Hawkins in the preparation of this manuscript,
and a prompt and helpful review by our anonymous referee.}
This material is based upon work supported by the National
Science Foundation under grant No.\ 2009136. 
Support for KK was provided by NASA through the NASA Hubble Fellowship grant HST-HF2-51510 awarded by the Space Telescope Science Institute, which is operated by the Association of Universities for Research in Astronomy, Inc., for NASA, under contract NAS5-26555.
This publication makes use of data products
from the Wide-field Infrared Survey Explorer, which is a joint
project of the University of California, Los Angeles, and the Jet
Propulsion Laboratory/California Institute of Technology, and
NEOWISE, which is a project of the Jet Propulsion
Laboratory/California Institute of Technology. WISE and
NEOWISE are funded by the National Aeronautics and Space
Administration. 
This research has benefited from the SpeX Prism Libraries Analysis Toolkit, maintained by Adam Burgasser at \url{https://github.com/aburgasser/splat}.
This research has made use of the SIMBAD database \citep{2000AAS..143....9W},
the ``Aladin sky atlas'' \citep{2000AAS..143...33B},
and the VizieR catalogue access tool
developed and operated at CDS, Strasbourg, France
Data presented herein were
obtained at Keck Observatory, which is a private 501(c)3
nonprofit organization operated as a scientific partnership among
the California Institute of Technology, the University of
California, and the National Aeronautics and Space Administration. The Observatory was made possible by the generous
financial support of the W. M. Keck Foundation. 
The authors recognize and acknowledge the significant cultural role
and reverence that the summit of Maunakea has within the
indigenous Hawaiian community, and that the W. M. Keck
Observatory stands on Crown and Government Lands that the
State of Hawai’i is obligated to protect and preserve for future
generations of indigenous Hawaiians. 
Portions of this work
were conducted at the University of California, San Diego,
which was built on the unceded territory of the Kumeyaay
Nation, whose people continue to maintain their political
sovereignty and cultural traditions as vital members of the San
Diego community.
\end{acknowledgments}

\vspace{5mm}
\facilities{Keck Observatory(NIRES)}

\software{
astropy \citep{2013A&A...558A..33A,2018AJ....156..123A,2022ApJ...935..167A},  
galpy \citep{2015ApJS..216...29B},
Matplotlib \citep{2007CSE.....9...90H},
NumPy \citep{2011CSE....13b..22V},
pandas \citep{mckinney-proc-scipy-2010},
SciPy \citep{2020NatMe..17..261V},
SpeXTool \citep{2004PASP..116..362C},
SPLAT \citep{2017ASInC..14....7B}
}

\bibliography{bib}{}

\begin{thebibliography}{}
\expandafter\ifx\csname natexlab\endcsname\relax\def\natexlab#1{#1}\fi
\providecommand{\url}[1]{\href{#1}{#1}}
\providecommand{\dodoi}[1]{doi:~\href{http://doi.org/#1}{\nolinkurl{#1}}}
\providecommand{\doeprint}[1]{\href{http://ascl.net/#1}{\nolinkurl{http://ascl.net/#1}}}
\providecommand{\doarXiv}[1]{\href{https://arxiv.org/abs/#1}{\nolinkurl{https://arxiv.org/abs/#1}}}

\bibitem[{{Abadi} {et~al.}(2009){Abadi}, {Navarro}, \& {Steinmetz}}]{2009ApJ...691L..63A}
{Abadi}, M.~G., {Navarro}, J.~F., \& {Steinmetz}, M. 2009, \apjl, 691, L63, \dodoi{10.1088/0004-637X/691/2/L63}

\bibitem[{{Allard} {et~al.}(2012){Allard}, {Homeier}, \& {Freytag}}]{2012RSPTA.370.2765A}
{Allard}, F., {Homeier}, D., \& {Freytag}, B. 2012, Philosophical Transactions of the Royal Society A, 370, 2765, \dodoi{10.1098/rsta.2011.0269}

\bibitem[{Alvarado {et~al.}(2024)Alvarado, Gerasimov, Burgasser, Brooks, Aganze, \& Theissen}]{Alvarado_2024}
Alvarado, E., Gerasimov, R., Burgasser, A.~J., {et~al.} 2024, Research Notes of the AAS, 8, 134, \dodoi{10.3847/2515-5172/ad4bd7}

\bibitem[{{Astropy Collaboration} {et~al.}(2013){Astropy Collaboration}, {Robitaille}, {Tollerud}, {Greenfield}, {Droettboom}, {Bray}, {Aldcroft}, {Davis}, {Ginsburg}, {Price-Whelan}, {Kerzendorf}, {Conley}, {Crighton}, {Barbary}, {Muna}, {Ferguson}, {Grollier}, {Parikh}, {Nair}, {Unther}, {Deil}, {Woillez}, {Conseil}, {Kramer}, {Turner}, {Singer}, {Fox}, {Weaver}, {Zabalza}, {Edwards}, {Azalee Bostroem}, {Burke}, {Casey}, {Crawford}, {Dencheva}, {Ely}, {Jenness}, {Labrie}, {Lim}, {Pierfederici}, {Pontzen}, {Ptak}, {Refsdal}, {Servillat}, \& {Streicher}}]{2013A&A...558A..33A}
{Astropy Collaboration}, {Robitaille}, T.~P., {Tollerud}, E.~J., {et~al.} 2013, \aap, 558, A33, \dodoi{10.1051/0004-6361/201322068}

\bibitem[{{Astropy Collaboration} {et~al.}(2018){Astropy Collaboration}, {Price-Whelan}, {Sip{\H{o}}cz}, {G{\"u}nther}, {Lim}, {Crawford}, {Conseil}, {Shupe}, {Craig}, {Dencheva}, {Ginsburg}, {VanderPlas}, {Bradley}, {P{\'e}rez-Su{\'a}rez}, {de Val-Borro}, {Aldcroft}, {Cruz}, {Robitaille}, {Tollerud}, {Ardelean}, {Babej}, {Bach}, {Bachetti}, {Bakanov}, {Bamford}, {Barentsen}, {Barmby}, {Baumbach}, {Berry}, {Biscani}, {Boquien}, {Bostroem}, {Bouma}, {Brammer}, {Bray}, {Breytenbach}, {Buddelmeijer}, {Burke}, {Calderone}, {Cano Rodr{\'\i}guez}, {Cara}, {Cardoso}, {Cheedella}, {Copin}, {Corrales}, {Crichton}, {D'Avella}, {Deil}, {Depagne}, {Dietrich}, {Donath}, {Droettboom}, {Earl}, {Erben}, {Fabbro}, {Ferreira}, {Finethy}, {Fox}, {Garrison}, {Gibbons}, {Goldstein}, {Gommers}, {Greco}, {Greenfield}, {Groener}, {Grollier}, {Hagen}, {Hirst}, {Homeier}, {Horton}, {Hosseinzadeh}, {Hu}, {Hunkeler}, {Ivezi{\'c}}, {Jain}, {Jenness}, {Kanarek}, {Kendrew}, {Kern}, {Kerzendorf}, {Khvalko}, {King}, {Kirkby}, {Kulkarni},
  {Kumar}, {Lee}, {Lenz}, {Littlefair}, {Ma}, {Macleod}, {Mastropietro}, {McCully}, {Montagnac}, {Morris}, {Mueller}, {Mumford}, {Muna}, {Murphy}, {Nelson}, {Nguyen}, {Ninan}, {N{\"o}the}, {Ogaz}, {Oh}, {Parejko}, {Parley}, {Pascual}, {Patil}, {Patil}, {Plunkett}, {Prochaska}, {Rastogi}, {Reddy Janga}, {Sabater}, {Sakurikar}, {Seifert}, {Sherbert}, {Sherwood-Taylor}, {Shih}, {Sick}, {Silbiger}, {Singanamalla}, {Singer}, {Sladen}, {Sooley}, {Sornarajah}, {Streicher}, {Teuben}, {Thomas}, {Tremblay}, {Turner}, {Terr{\'o}n}, {van Kerkwijk}, {de la Vega}, {Watkins}, {Weaver}, {Whitmore}, {Woillez}, {Zabalza}, \& {Astropy Contributors}}]{2018AJ....156..123A}
{Astropy Collaboration}, {Price-Whelan}, A.~M., {Sip{\H{o}}cz}, B.~M., {et~al.} 2018, \aj, 156, 123, \dodoi{10.3847/1538-3881/aabc4f}

\bibitem[{{Astropy Collaboration} {et~al.}(2022){Astropy Collaboration}, {Price-Whelan}, {Lim}, {Earl}, {Starkman}, {Bradley}, {Shupe}, {Patil}, {Corrales}, {Brasseur}, {N{\"o}the}, {Donath}, {Tollerud}, {Morris}, {Ginsburg}, {Vaher}, {Weaver}, {Tocknell}, {Jamieson}, {van Kerkwijk}, {Robitaille}, {Merry}, {Bachetti}, {G{\"u}nther}, {Aldcroft}, {Alvarado-Montes}, {Archibald}, {B{\'o}di}, {Bapat}, {Barentsen}, {Baz{\'a}n}, {Biswas}, {Boquien}, {Burke}, {Cara}, {Cara}, {Conroy}, {Conseil}, {Craig}, {Cross}, {Cruz}, {D'Eugenio}, {Dencheva}, {Devillepoix}, {Dietrich}, {Eigenbrot}, {Erben}, {Ferreira}, {Foreman-Mackey}, {Fox}, {Freij}, {Garg}, {Geda}, {Glattly}, {Gondhalekar}, {Gordon}, {Grant}, {Greenfield}, {Groener}, {Guest}, {Gurovich}, {Handberg}, {Hart}, {Hatfield-Dodds}, {Homeier}, {Hosseinzadeh}, {Jenness}, {Jones}, {Joseph}, {Kalmbach}, {Karamehmetoglu}, {Ka{\l}uszy{\'n}ski}, {Kelley}, {Kern}, {Kerzendorf}, {Koch}, {Kulumani}, {Lee}, {Ly}, {Ma}, {MacBride}, {Maljaars}, {Muna}, {Murphy}, {Norman},
  {O'Steen}, {Oman}, {Pacifici}, {Pascual}, {Pascual-Granado}, {Patil}, {Perren}, {Pickering}, {Rastogi}, {Roulston}, {Ryan}, {Rykoff}, {Sabater}, {Sakurikar}, {Salgado}, {Sanghi}, {Saunders}, {Savchenko}, {Schwardt}, {Seifert-Eckert}, {Shih}, {Jain}, {Shukla}, {Sick}, {Simpson}, {Singanamalla}, {Singer}, {Singhal}, {Sinha}, {Sip{\H{o}}cz}, {Spitler}, {Stansby}, {Streicher}, {{\v{S}}umak}, {Swinbank}, {Taranu}, {Tewary}, {Tremblay}, {de Val-Borro}, {Van Kooten}, {Vasovi{\'c}}, {Verma}, {de Miranda Cardoso}, {Williams}, {Wilson}, {Winkel}, {Wood-Vasey}, {Xue}, {Yoachim}, {Zhang}, {Zonca}, \& {Astropy Project Contributors}}]{2022ApJ...935..167A}
{Astropy Collaboration}, {Price-Whelan}, A.~M., {Lim}, P.~L., {et~al.} 2022, \apj, 935, 167, \dodoi{10.3847/1538-4357/ac7c74}

\bibitem[{{Bardalez Gagliuffi} {et~al.}(2014){Bardalez Gagliuffi}, {Burgasser}, {Gelino}, {Looper}, {Nicholls}, {Schmidt}, {Cruz}, {West}, {Gizis}, \& {Metchev}}]{2014ApJ...794..143B}
{Bardalez Gagliuffi}, D.~C., {Burgasser}, A.~J., {Gelino}, C.~R., {et~al.} 2014, \apj, 794, 143, \dodoi{10.1088/0004-637X/794/2/143}

\bibitem[{{Blaauw}(1961)}]{1961BAN....15..265B}
{Blaauw}, A. 1961, \bain, 15, 265

\bibitem[{{Blake} {et~al.}(2010){Blake}, {Charbonneau}, \& {White}}]{2010ApJ...723..684B}
{Blake}, C.~H., {Charbonneau}, D., \& {White}, R.~J. 2010, \apj, 723, 684, \dodoi{10.1088/0004-637X/723/1/684}

\bibitem[{{Bonnarel} {et~al.}(2000){Bonnarel}, {Fernique}, {Bienaym{\'e}}, {Egret}, {Genova}, {Louys}, {Ochsenbein}, {Wenger}, \& {Bartlett}}]{2000AAS..143...33B}
{Bonnarel}, F., {Fernique}, P., {Bienaym{\'e}}, O., {et~al.} 2000, \aaps, 143, 33, \dodoi{10.1051/aas:2000331}

\bibitem[{{Bovy}(2015)}]{2015ApJS..216...29B}
{Bovy}, J. 2015, \apjs, 216, 29, \dodoi{10.1088/0067-0049/216/2/29}

\bibitem[{{Brooks} {et~al.}(2022){Brooks}, {Kirkpatrick}, {Caselden}, {Schneider}, {Meisner}, {Faherty}, {Casewell}, {Kuchner}, {Kuchner}, \& {Backyard Worlds: Planet 9 Collaboration}}]{2022AJ....163...47B}
{Brooks}, H., {Kirkpatrick}, J.~D., {Caselden}, D., {et~al.} 2022, \aj, 163, 47, \dodoi{10.3847/1538-3881/ac3a0a}

\bibitem[{{Brown}(2015)}]{2015ARA&A..53...15B}
{Brown}, W.~R. 2015, \araa, 53, 15, \dodoi{10.1146/annurev-astro-082214-122230}

\bibitem[{{Burgasser} {et~al.}(2007){Burgasser}, {Cruz}, \& {Kirkpatrick}}]{2007ApJ...657..494B}
{Burgasser}, A.~J., {Cruz}, K.~L., \& {Kirkpatrick}, J.~D. 2007, \apj, 657, 494, \dodoi{10.1086/510148}

\bibitem[{{Burgasser} \& {Splat Development Team}(2017)}]{2017ASInC..14....7B}
{Burgasser}, A.~J., \& {Splat Development Team}. 2017, in Astronomical Society of India Conference Series, Vol.~14, Astronomical Society of India Conference Series, 7--12.
\newblock \doarXiv{1707.00062}

\bibitem[{{Burgasser} {et~al.}(2003){Burgasser}, {Kirkpatrick}, {Burrows}, {Liebert}, {Reid}, {Gizis}, {McGovern}, {Prato}, \& {McLean}}]{2003ApJ...592.1186B}
{Burgasser}, A.~J., {Kirkpatrick}, J.~D., {Burrows}, A., {et~al.} 2003, \apj, 592, 1186, \dodoi{10.1086/375813}

\bibitem[{{Burgasser} {et~al.}(2015){Burgasser}, {Gillon}, {Melis}, {Bowler}, {Michelsen}, {Bardalez Gagliuffi}, {Gelino}, {Jehin}, {Delrez}, {Manfroid}, \& {Blake}}]{2015AJ....149..104B}
{Burgasser}, A.~J., {Gillon}, M., {Melis}, C., {et~al.} 2015, \aj, 149, 104, \dodoi{10.1088/0004-6256/149/3/104}

\bibitem[{{Burgasser} {et~al.}(2024){Burgasser}, {Bruursema}, {Munn}, {Vrba}, {Caselden}, {Kabatnik}, {Rothermich}, {Sainio}, {Bickle}, {Dahm}, {Meisner}, {Kirkpatrick}, {Su{\'a}rez}, {Gagn{\'e}}, {Faherty}, {Vos}, {Kuchner}, {Williams}, {Bardalez Gagliuffi}, {Aganze}, {Hsu}, {Theissen}, {Cushing}, {Marocco}, {Casewell}, \& {Backyard Worlds: Planet 9 Collaboration}}]{burgasser2024}
{Burgasser}, A.~J., {Bruursema}, J., {Munn}, J.~A., {et~al.} 2024, \apj, in review

\bibitem[{{Cabrera} \& {Rodriguez}(2023)}]{2023ApJ...953...19C}
{Cabrera}, T., \& {Rodriguez}, C.~L. 2023, \apj, 953, 19, \dodoi{10.3847/1538-4357/acdc22}

\bibitem[{{Carr} {et~al.}(2000){Carr}, {Sellgren}, \& {Balachandran}}]{2000ApJ...530..307C}
{Carr}, J.~S., {Sellgren}, K., \& {Balachandran}, S.~C. 2000, \apj, 530, 307, \dodoi{10.1086/308340}

\bibitem[{{Chambers} {et~al.}(2016){Chambers}, {Magnier}, {Metcalfe}, {Flewelling}, {Huber}, {Waters}, {Denneau}, {Draper}, {Farrow}, {Finkbeiner}, {Holmberg}, {Koppenhoefer}, {Price}, {Rest}, {Saglia}, {Schlafly}, {Smartt}, {Sweeney}, {Wainscoat}, {Burgett}, {Chastel}, {Grav}, {Heasley}, {Hodapp}, {Jedicke}, {Kaiser}, {Kudritzki}, {Luppino}, {Lupton}, {Monet}, {Morgan}, {Onaka}, {Shiao}, {Stubbs}, {Tonry}, {White}, {Ba{\~n}ados}, {Bell}, {Bender}, {Bernard}, {Boegner}, {Boffi}, {Botticella}, {Calamida}, {Casertano}, {Chen}, {Chen}, {Cole}, {Deacon}, {Frenk}, {Fitzsimmons}, {Gezari}, {Gibbs}, {Goessl}, {Goggia}, {Gourgue}, {Goldman}, {Grant}, {Grebel}, {Hambly}, {Hasinger}, {Heavens}, {Heckman}, {Henderson}, {Henning}, {Holman}, {Hopp}, {Ip}, {Isani}, {Jackson}, {Keyes}, {Koekemoer}, {Kotak}, {Le}, {Liska}, {Long}, {Lucey}, {Liu}, {Martin}, {Masci}, {McLean}, {Mindel}, {Misra}, {Morganson}, {Murphy}, {Obaika}, {Narayan}, {Nieto-Santisteban}, {Norberg}, {Peacock}, {Pier}, {Postman}, {Primak}, {Rae}, {Rai},
  {Riess}, {Riffeser}, {Rix}, {R{\"o}ser}, {Russel}, {Rutz}, {Schilbach}, {Schultz}, {Scolnic}, {Strolger}, {Szalay}, {Seitz}, {Small}, {Smith}, {Soderblom}, {Taylor}, {Thomson}, {Taylor}, {Thakar}, {Thiel}, {Thilker}, {Unger}, {Urata}, {Valenti}, {Wagner}, {Walder}, {Walter}, {Watters}, {Werner}, {Wood-Vasey}, \& {Wyse}}]{2016arXiv161205560C}
{Chambers}, K.~C., {Magnier}, E.~A., {Metcalfe}, N., {et~al.} 2016, arXiv e-prints, arXiv:1612.05560, \dodoi{10.48550/arXiv.1612.05560}

\bibitem[{{Chandrasekhar}(1931)}]{1931MNRAS..91..456C}
{Chandrasekhar}, S. 1931, \mnras, 91, 456, \dodoi{10.1093/mnras/91.5.456}

\bibitem[{{Chandrasekhar}(1935)}]{1935MNRAS..95..207C}
---. 1935, \mnras, 95, 207, \dodoi{10.1093/mnras/95.3.207}

\bibitem[{{Cui} {et~al.}(2012){Cui}, {Zhao}, {Chu}, {Li}, {Li}, {Zhang}, {Su}, {Yao}, {Wang}, {Xing}, {Li}, {Zhu}, {Wang}, {Gu}, {Luo}, {Xu}, {Zhang}, {Liu}, {Zhang}, {Yang}, {Cao}, {Chen}, {Chen}, {Chen}, {Chen}, {Chu}, {Feng}, {Gong}, {Hou}, {Hu}, {Hu}, {Hu}, {Jia}, {Jiang}, {Jiang}, {Jiang}, {Jin}, {Li}, {Li}, {Li}, {Liu}, {Liu}, {Lu}, {Mao}, {Men}, {Qi}, {Qi}, {Shi}, {Tang}, {Tao}, {Wang}, {Wang}, {Wang}, {Wang}, {Wang}, {Wang}, {Wang}, {Wang}, {Wang}, {Wang}, {Wang}, {Wang}, {Xu}, {Xu}, {Yang}, {Yu}, {Yuan}, {Yuan}, {Zhai}, {Zhang}, {Zhang}, {Zhang}, {Zhao}, {Zhou}, {Zhou}, {Zhu}, \& {Zou}}]{2012RAA....12.1197C}
{Cui}, X.-Q., {Zhao}, Y.-H., {Chu}, Y.-Q., {et~al.} 2012, Research in Astronomy and Astrophysics, 12, 1197, \dodoi{10.1088/1674-4527/12/9/003}

\bibitem[{{Cunha} {et~al.}(2007){Cunha}, {Sellgren}, {Smith}, {Ramirez}, {Blum}, \& {Terndrup}}]{2007ApJ...669.1011C}
{Cunha}, K., {Sellgren}, K., {Smith}, V.~V., {et~al.} 2007, \apj, 669, 1011, \dodoi{10.1086/521813}

\bibitem[{{Cushing} {et~al.}(2004){Cushing}, {Vacca}, \& {Rayner}}]{2004PASP..116..362C}
{Cushing}, M.~C., {Vacca}, W.~D., \& {Rayner}, J.~T. 2004, \pasp, 116, 362, \dodoi{10.1086/382907}

\bibitem[{{Daniel} \& {Wyse}(2018)}]{2018MNRAS.476.1561D}
{Daniel}, K.~J., \& {Wyse}, R. F.~G. 2018, \mnras, 476, 1561, \dodoi{10.1093/mnras/sty199}

\bibitem[{{Drlica-Wagner} {et~al.}(2015){Drlica-Wagner}, {Bechtol}, {Rykoff}, {Luque}, {Queiroz}, {Mao}, {Wechsler}, {Simon}, {Santiago}, {Yanny}, {Balbinot}, {Dodelson}, {Fausti Neto}, {James}, {Li}, {Maia}, {Marshall}, {Pieres}, {Stringer}, {Walker}, {Abbott}, {Abdalla}, {Allam}, {Benoit-L{\'e}vy}, {Bernstein}, {Bertin}, {Brooks}, {Buckley-Geer}, {Burke}, {Carnero Rosell}, {Carrasco Kind}, {Carretero}, {Crocce}, {da Costa}, {Desai}, {Diehl}, {Dietrich}, {Doel}, {Eifler}, {Evrard}, {Finley}, {Flaugher}, {Fosalba}, {Frieman}, {Gaztanaga}, {Gerdes}, {Gruen}, {Gruendl}, {Gutierrez}, {Honscheid}, {Kuehn}, {Kuropatkin}, {Lahav}, {Martini}, {Miquel}, {Nord}, {Ogando}, {Plazas}, {Reil}, {Roodman}, {Sako}, {Sanchez}, {Scarpine}, {Schubnell}, {Sevilla-Noarbe}, {Smith}, {Soares-Santos}, {Sobreira}, {Suchyta}, {Swanson}, {Tarle}, {Tucker}, {Vikram}, {Wester}, {Zhang}, {Zuntz}, \& {DES Collaboration}}]{2015ApJ...813..109D}
{Drlica-Wagner}, A., {Bechtol}, K., {Rykoff}, E.~S., {et~al.} 2015, \apj, 813, 109, \dodoi{10.1088/0004-637X/813/2/109}

\bibitem[{{Du} {et~al.}(2019){Du}, {Li}, {Yan}, {Newberg}, {Shi}, {Ma}, {Chen}, \& {Wu}}]{2019ApJS..244....4D}
{Du}, C., {Li}, H., {Yan}, Y., {et~al.} 2019, \apjs, 244, 4, \dodoi{10.3847/1538-4365/ab328c}

\bibitem[{{Dye} {et~al.}(2018){Dye}, {Lawrence}, {Read}, {Fan}, {Kerr}, {Varricatt}, {Furnell}, {Edge}, {Irwin}, {Hambly}, {Lucas}, {Almaini}, {Chambers}, {Green}, {Hewett}, {Liu}, {McGreer}, {Best}, {Zhang}, {Sutorius}, {Froebrich}, {Magnier}, {Hasinger}, {Lederer}, {Bold}, \& {Tedds}}]{2018MNRAS.473.5113D}
{Dye}, S., {Lawrence}, A., {Read}, M.~A., {et~al.} 2018, \mnras, 473, 5113, \dodoi{10.1093/mnras/stx2622}

\bibitem[{{Eggleton}(1983)}]{1983ApJ...268..368E}
{Eggleton}, P.~P. 1983, \apj, 268, 368, \dodoi{10.1086/160960}

\bibitem[{{Favia} {et~al.}(2015){Favia}, {West}, \& {Theissen}}]{2015ApJ...813...26F}
{Favia}, A., {West}, A.~A., \& {Theissen}, C.~A. 2015, \apj, 813, 26, \dodoi{10.1088/0004-637X/813/1/26}

\bibitem[{{Fragione} \& {Gualandris}(2019)}]{2019MNRAS.489.4543F}
{Fragione}, G., \& {Gualandris}, A. 2019, \mnras, 489, 4543, \dodoi{10.1093/mnras/stz2451}

\bibitem[{{Fregeau} {et~al.}(2002){Fregeau}, {Joshi}, {Portegies Zwart}, \& {Rasio}}]{2002ApJ...570..171F}
{Fregeau}, J.~M., {Joshi}, K.~J., {Portegies Zwart}, S.~F., \& {Rasio}, F.~A. 2002, \apj, 570, 171, \dodoi{10.1086/339576}

\bibitem[{{Fritz} {et~al.}(2018){Fritz}, {Battaglia}, {Pawlowski}, {Kallivayalil}, {van der Marel}, {Sohn}, {Brook}, \& {Besla}}]{2018A&A...619A.103F}
{Fritz}, T.~K., {Battaglia}, G., {Pawlowski}, M.~S., {et~al.} 2018, \aap, 619, A103, \dodoi{10.1051/0004-6361/201833343}

\bibitem[{{Gaia Collaboration} {et~al.}(2021){Gaia Collaboration}, {Brown}, {Vallenari}, {Prusti}, {de Bruijne}, {Babusiaux}, {Biermann}, {Creevey}, {Evans}, {Eyer}, \& et~al.}]{2021A&A...650C...3G}
{Gaia Collaboration}, {Brown}, A.~G.~A., {Vallenari}, A., {et~al.} 2021, \aap, 650, C3, \dodoi{10.1051/0004-6361/202039657e}

\bibitem[{{Gerasimov} {et~al.}(2024){Gerasimov}, {Bedin}, {Burgasser}, {Apai}, {Nardiello}, {Alvarado}, \& {Anderson}}]{2024arXiv240501634G}
{Gerasimov}, R., {Bedin}, L.~R., {Burgasser}, A.~J., {et~al.} 2024, arXiv e-prints, arXiv:2405.01634, \dodoi{10.48550/arXiv.2405.01634}

\bibitem[{{Gerasimov} {et~al.}(2022){Gerasimov}, {Burgasser}, {Homeier}, {Bedin}, {Rees}, {Scalco}, {Anderson}, \& {Salaris}}]{2022ApJ...930...24G}
{Gerasimov}, R., {Burgasser}, A.~J., {Homeier}, D., {et~al.} 2022, \apj, 930, 24, \dodoi{10.3847/1538-4357/ac61e5}

\bibitem[{{Gonzales} {et~al.}(2021){Gonzales}, {Burningham}, {Faherty}, {Visscher}, {Marley}, {Lupu}, {Freedman}, \& {Lewis}}]{2021ApJ...923...19G}
{Gonzales}, E.~C., {Burningham}, B., {Faherty}, J.~K., {et~al.} 2021, \apj, 923, 19, \dodoi{10.3847/1538-4357/ac294e}

\bibitem[{{Gonzales} {et~al.}(2018){Gonzales}, {Faherty}, {Gagn{\'e}}, {Artigau}, \& {Bardalez Gagliuffi}}]{2018ApJ...864..100G}
{Gonzales}, E.~C., {Faherty}, J.~K., {Gagn{\'e}}, J., {Artigau}, {\'E}., \& {Bardalez Gagliuffi}, D. 2018, \apj, 864, 100, \dodoi{10.3847/1538-4357/aad3c7}

\bibitem[{{Harris}(1996)}]{1996AJ....112.1487H}
{Harris}, W.~E. 1996, \aj, 112, 1487, \dodoi{10.1086/118116}

\bibitem[{{Hastings}(1970)}]{HASTINGS01041970}
{Hastings}, W.~K. 1970, Biometrika, 57, 97, \dodoi{10.1093/biomet/57.1.97}

\bibitem[{{Hawkins} \& {Wyse}(2018)}]{2018MNRAS.481.1028H}
{Hawkins}, K., \& {Wyse}, R. F.~G. 2018, \mnras, 481, 1028, \dodoi{10.1093/mnras/sty2282}

\bibitem[{{Herschel}(1864)}]{1864RSPT..154....1H}
{Herschel}, J. F.~W. 1864, Philosophical Transactions of the Royal Society of London Series I, 154, 1

\bibitem[{{Hills}(1988)}]{1988Natur.331..687H}
{Hills}, J.~G. 1988, \nat, 331, 687, \dodoi{10.1038/331687a0}

\bibitem[{{Hunter}(2007)}]{2007CSE.....9...90H}
{Hunter}, J.~D. 2007, Computing in Science and Engineering, 9, 90, \dodoi{10.1109/MCSE.2007.55}

\bibitem[{{Kirkpatrick} {et~al.}(2010){Kirkpatrick}, {Looper}, {Burgasser}, {Schurr}, {Cutri}, {Cushing}, {Cruz}, {Sweet}, {Knapp}, {Barman}, {Bochanski}, {Roellig}, {McLean}, {McGovern}, \& {Rice}}]{2010ApJS..190..100K}
{Kirkpatrick}, J.~D., {Looper}, D.~L., {Burgasser}, A.~J., {et~al.} 2010, \apjs, 190, 100, \dodoi{10.1088/0067-0049/190/1/100}

\bibitem[{{Kirkpatrick} {et~al.}(2021{\natexlab{a}}){Kirkpatrick}, {Marocco}, {Caselden}, {Meisner}, {Faherty}, {Schneider}, {Kuchner}, {Casewell}, {Gelino}, {Cushing}, {Eisenhardt}, {Wright}, \& {Schurr}}]{2021ApJ...915L...6K}
{Kirkpatrick}, J.~D., {Marocco}, F., {Caselden}, D., {et~al.} 2021{\natexlab{a}}, \apjl, 915, L6, \dodoi{10.3847/2041-8213/ac0437}

\bibitem[{{Kirkpatrick} {et~al.}(2021{\natexlab{b}}){Kirkpatrick}, {Gelino}, {Faherty}, {Meisner}, {Caselden}, {Schneider}, {Marocco}, {Cayago}, {Smart}, {Eisenhardt}, {Kuchner}, {Wright}, {Cushing}, {Allers}, {Bardalez Gagliuffi}, {Burgasser}, {Gagn{\'e}}, {Logsdon}, {Martin}, {Ingalls}, {Lowrance}, {Abrahams}, {Aganze}, {Gerasimov}, {Gonzales}, {Hsu}, {Kamraj}, {Kiman}, {Rees}, {Theissen}, {Ammar}, {Andersen}, {Beaulieu}, {Colin}, {Elachi}, {Goodman}, {Gramaize}, {Hamlet}, {Hong}, {Jonkeren}, {Khalil}, {Martin}, {Pendrill}, {Pumphrey}, {Rothermich}, {Sainio}, {Stenner}, {Tanner}, {Th{\'e}venot}, {Voloshin}, {Walla}, {W{\k{e}}dracki}, \& {Backyard Worlds: Planet 9 Collaboration}}]{2021ApJS..253....7K}
{Kirkpatrick}, J.~D., {Gelino}, C.~R., {Faherty}, J.~K., {et~al.} 2021{\natexlab{b}}, \apjs, 253, 7, \dodoi{10.3847/1538-4365/abd107}

\bibitem[{{Kolb} \& {Baraffe}(1999)}]{1999MNRAS.309.1034K}
{Kolb}, U., \& {Baraffe}, I. 1999, \mnras, 309, 1034, \dodoi{10.1046/j.1365-8711.1999.02926.x}

\bibitem[{{Konopacky} {et~al.}(2010){Konopacky}, {Ghez}, {Barman}, {Rice}, {Bailey}, {White}, {McLean}, \& {Duch{\^e}ne}}]{2010ApJ...711.1087K}
{Konopacky}, Q.~M., {Ghez}, A.~M., {Barman}, T.~S., {et~al.} 2010, \apj, 711, 1087, \dodoi{10.1088/0004-637X/711/2/1087}

\bibitem[{{Kremer} {et~al.}(2020{\natexlab{a}}){Kremer}, {Ye}, {Chatterjee}, {Rodriguez}, \& {Rasio}}]{2020IAUS..351..357K}
{Kremer}, K., {Ye}, C.~S., {Chatterjee}, S., {Rodriguez}, C.~L., \& {Rasio}, F.~A. 2020{\natexlab{a}}, in Star Clusters: From the Milky Way to the Early Universe, ed. A.~{Bragaglia}, M.~{Davies}, A.~{Sills}, \& E.~{Vesperini}, Vol. 351, 357--366, \dodoi{10.1017/S1743921319007269}

\bibitem[{{Kremer} {et~al.}(2020{\natexlab{b}}){Kremer}, {Ye}, {Rui}, {Weatherford}, {Chatterjee}, {Fragione}, {Rodriguez}, {Spera}, \& {Rasio}}]{2020ApJS..247...48K}
{Kremer}, K., {Ye}, C.~S., {Rui}, N.~Z., {et~al.} 2020{\natexlab{b}}, \apjs, 247, 48, \dodoi{10.3847/1538-4365/ab7919}

\bibitem[{{Kroupa}(2001)}]{2001MNRAS.322..231K}
{Kroupa}, P. 2001, \mnras, 322, 231, \dodoi{10.1046/j.1365-8711.2001.04022.x}

\bibitem[{{Kuchner} {et~al.}(2017){Kuchner}, {Faherty}, {Schneider}, {Meisner}, {Filippazzo}, {Gagn{\'e}}, {Trouille}, {Silverberg}, {Castro}, {Fletcher}, {Mokaev}, \& {Stajic}}]{2017ApJ...841L..19K}
{Kuchner}, M.~J., {Faherty}, J.~K., {Schneider}, A.~C., {et~al.} 2017, \apjl, 841, L19, \dodoi{10.3847/2041-8213/aa7200}

\bibitem[{{Lang}(2014)}]{2014AJ....147..108L}
{Lang}, D. 2014, \aj, 147, 108, \dodoi{10.1088/0004-6256/147/5/108}

\bibitem[{{Leahy} \& {Williams}(2017)}]{2017AJ....153..239L}
{Leahy}, D.~A., \& {Williams}, J.~E. 2017, \aj, 153, 239, \dodoi{10.3847/1538-3881/aa6af6}

\bibitem[{{Leonard}(1991)}]{1991AJ....101..562L}
{Leonard}, P. J.~T. 1991, \aj, 101, 562, \dodoi{10.1086/115704}

\bibitem[{{Li} {et~al.}(2023){Li}, {Huang}, {Dong}, {Zhang}, {Beers}, \& {Yuan}}]{2023AJ....166...12L}
{Li}, Q.-Z., {Huang}, Y., {Dong}, X.-B., {et~al.} 2023, \aj, 166, 12, \dodoi{10.3847/1538-3881/acd1dc}

\bibitem[{{Li} {et~al.}(2012){Li}, {Luo}, {Zhao}, {Lu}, {Ren}, \& {Zuo}}]{2012ApJ...744L..24L}
{Li}, Y., {Luo}, A., {Zhao}, G., {et~al.} 2012, \apjl, 744, L24, \dodoi{10.1088/2041-8205/744/2/L24}

\bibitem[{{Liao} {et~al.}(2023){Liao}, {Du}, {Li}, {Ma}, \& {Shi}}]{2023ApJ...944L..39L}
{Liao}, J., {Du}, C., {Li}, H., {Ma}, J., \& {Shi}, J. 2023, \apjl, 944, L39, \dodoi{10.3847/2041-8213/acb7d9}

\bibitem[{{Linsky}(1969)}]{1969ApJ...156..989L}
{Linsky}, J.~L. 1969, \apj, 156, 989, \dodoi{10.1086/150030}

\bibitem[{{Littlefair} {et~al.}(2006){Littlefair}, {Dhillon}, {Marsh}, {G{\"a}nsicke}, {Southworth}, \& {Watson}}]{2006Sci...314.1578L}
{Littlefair}, S.~P., {Dhillon}, V.~S., {Marsh}, T.~R., {et~al.} 2006, Science, 314, 1578, \dodoi{10.1126/science.1133333}

\bibitem[{{Livingston} \& {Wallace}(1991)}]{1991aass.book.....L}
{Livingston}, W., \& {Wallace}, L. 1991, {An atlas of the solar spectrum in the infrared from 1850 to 9000 cm-1 (1.1 to 5.4 micrometer)}

\bibitem[{{Longstaff} {et~al.}(2019){Longstaff}, {Casewell}, {Wynn}, {Page}, {Williams}, {Braker}, \& {Maxted}}]{2019MNRAS.484.2566L}
{Longstaff}, E.~S., {Casewell}, S.~L., {Wynn}, G.~A., {et~al.} 2019, \mnras, 484, 2566, \dodoi{10.1093/mnras/stz127}

\bibitem[{{Luhman} \& {Sheppard}(2014)}]{2014ApJ...787..126L}
{Luhman}, K.~L., \& {Sheppard}, S.~S. 2014, \apj, 787, 126, \dodoi{10.1088/0004-637X/787/2/126}

\bibitem[{{Mainzer} {et~al.}(2014){Mainzer}, {Bauer}, {Cutri}, {Grav}, {Masiero}, {Beck}, {Clarkson}, {Conrow}, {Dailey}, {Eisenhardt}, {Fabinsky}, {Fajardo-Acosta}, {Fowler}, {Gelino}, {Grillmair}, {Heinrichsen}, {Kendall}, {Kirkpatrick}, {Liu}, {Masci}, {McCallon}, {Nugent}, {Papin}, {Rice}, {Royer}, {Ryan}, {Sevilla}, {Sonnett}, {Stevenson}, {Thompson}, {Wheelock}, {Wiemer}, {Wittman}, {Wright}, \& {Yan}}]{2014ApJ...792...30M}
{Mainzer}, A., {Bauer}, J., {Cutri}, R.~M., {et~al.} 2014, \apj, 792, 30, \dodoi{10.1088/0004-637X/792/1/30}

\bibitem[{{Majewski} {et~al.}(2017){Majewski}, {Schiavon}, {Frinchaboy}, {Allende Prieto}, {Barkhouser}, {Bizyaev}, {Blank}, {Brunner}, {Burton}, {Carrera}, {Chojnowski}, {Cunha}, {Epstein}, {Fitzgerald}, {Garc{\'{\i}}a P{\'e}rez}, {Hearty}, {Henderson}, {Holtzman}, {Johnson}, {Lam}, {Lawler}, {Maseman}, {M{\'e}sz{\'a}ros}, {Nelson}, {Nguyen}, {Nidever}, {Pinsonneault}, {Shetrone}, {Smee}, {Smith}, {Stolberg}, {Skrutskie}, {Walker}, {Wilson}, {Zasowski}, {Anders}, {Basu}, {Beland}, {Blanton}, {Bovy}, {Brownstein}, {Carlberg}, {Chaplin}, {Chiappini}, {Eisenstein}, {Elsworth}, {Feuillet}, {Fleming}, {Galbraith-Frew}, {Garc{\'{\i}}a}, {Garc{\'{\i}}a-Hern{\'a}ndez}, {Gillespie}, {Girardi}, {Gunn}, {Hasselquist}, {Hayden}, {Hekker}, {Ivans}, {Kinemuchi}, {Klaene}, {Mahadevan}, {Mathur}, {Mosser}, {Muna}, {Munn}, {Nichol}, {O'Connell}, {Parejko}, {Robin}, {Rocha-Pinto}, {Schultheis}, {Serenelli}, {Shane}, {Silva Aguirre}, {Sobeck}, {Thompson}, {Troup}, {Weinberg}, \& {Zamora}}]{2017AJ....154...94M}
{Majewski}, S.~R., {Schiavon}, R.~P., {Frinchaboy}, P.~M., {et~al.} 2017, \aj, 154, 94, \dodoi{10.3847/1538-3881/aa784d}

\bibitem[{{Marocco} {et~al.}(2021){Marocco}, {Eisenhardt}, {Fowler}, {Kirkpatrick}, {Meisner}, {Schlafly}, {Stanford}, {Garcia}, {Caselden}, {Cushing}, {Cutri}, {Faherty}, {Gelino}, {Gonzalez}, {Jarrett}, {Koontz}, {Mainzer}, {Marchese}, {Mobasher}, {Schlegel}, {Stern}, {Teplitz}, \& {Wright}}]{2021ApJS..253....8M}
{Marocco}, F., {Eisenhardt}, P. R.~M., {Fowler}, J.~W., {et~al.} 2021, \apjs, 253, 8, \dodoi{10.3847/1538-4365/abd805}

\bibitem[{{Meisner} {et~al.}(2018){Meisner}, {Lang}, \& {Schlegel}}]{2018AJ....156...69M}
{Meisner}, A.~M., {Lang}, D., \& {Schlegel}, D.~J. 2018, \aj, 156, 69, \dodoi{10.3847/1538-3881/aacbcd}

\bibitem[{{Meisner} {et~al.}(2020){Meisner}, {Faherty}, {Kirkpatrick}, {Schneider}, {Caselden}, {Gagn{\'e}}, {Kuchner}, {Burgasser}, {Casewell}, {Debes}, {Artigau}, {Bardalez Gagliuffi}, {Logsdon}, {Kiman}, {Allers}, {Hsu}, {Wisniewski}, {Allen}, {Beaulieu}, {Colin}, {Durantini Luca}, {Goodman}, {Gramaize}, {Hamlet}, {Hinckley}, {Kiwy}, {Martin}, {Pendrill}, {Rothermich}, {Sainio}, {Sch{\"u}mann}, {Andersen}, {Tanner}, {Thakur}, {Th{\'e}venot}, {Walla}, {W{\k{e}}dracki}, {Aganze}, {Gerasimov}, {Theissen}, \& {Backyard Worlds: Planet 9 Collaboration}}]{2020ApJ...899..123M}
{Meisner}, A.~M., {Faherty}, J.~K., {Kirkpatrick}, J.~D., {et~al.} 2020, \apj, 899, 123, \dodoi{10.3847/1538-4357/aba633}

\bibitem[{{Meisner} {et~al.}(2021){Meisner}, {Schneider}, {Burgasser}, {Marocco}, {Line}, {Faherty}, {Kirkpatrick}, {Caselden}, {Kuchner}, {Gelino}, {Gagn{\'e}}, {Theissen}, {Gerasimov}, {Aganze}, {Hsu}, {Wisniewski}, {Casewell}, {Bardalez Gagliuffi}, {Logsdon}, {Eisenhardt}, {Allers}, {Debes}, {Allen}, {Stevnbak Andersen}, {Goodman}, {Gramaize}, {Martin}, {Sainio}, {Cushing}, \& {Backyard Worlds: Planet 9 Collaboration}}]{2021ApJ...915..120M}
{Meisner}, A.~M., {Schneider}, A.~C., {Burgasser}, A.~J., {et~al.} 2021, \apj, 915, 120, \dodoi{10.3847/1538-4357/ac013c}

\bibitem[{{Metropolis} {et~al.}(1953){Metropolis}, {Rosenbluth}, {Rosenbluth}, {Teller}, \& {Teller}}]{1953JChPh..21.1087M}
{Metropolis}, N., {Rosenbluth}, A.~W., {Rosenbluth}, M.~N., {Teller}, A.~H., \& {Teller}, E. 1953, \jcp, 21, 1087, \dodoi{10.1063/1.1699114}

\bibitem[{{Monari} {et~al.}(2018){Monari}, {Famaey}, {Carrillo}, {Piffl}, {Steinmetz}, {Wyse}, {Anders}, {Chiappini}, \& {Jan{\ss}en}}]{2018A&A...616L...9M}
{Monari}, G., {Famaey}, B., {Carrillo}, I., {et~al.} 2018, \aap, 616, L9, \dodoi{10.1051/0004-6361/201833748}

\bibitem[{{Mukherjee} {et~al.}(2024){Mukherjee}, {Fortney}, {Morley}, {Batalha}, {Marley}, {Karalidi}, {Visscher}, {Lupu}, {Freedman}, \& {Gharib-Nezhad}}]{2024ApJ...963...73M}
{Mukherjee}, S., {Fortney}, J.~J., {Morley}, C.~V., {et~al.} 2024, \apj, 963, 73, \dodoi{10.3847/1538-4357/ad18c2}

\bibitem[{{Nelder}(1965)}]{nelder65}
{Nelder}, J.~A. \&~{Mead}, R. 1965, Computer Journal, 7, 308

\bibitem[{{Pan} {et~al.}(2012){Pan}, {Ricker}, \& {Taam}}]{2012ApJ...750..151P}
{Pan}, K.-C., {Ricker}, P.~M., \& {Taam}, R.~E. 2012, \apj, 750, 151, \dodoi{10.1088/0004-637X/750/2/151}

\bibitem[{{Piffl} {et~al.}(2011){Piffl}, {Williams}, \& {Steinmetz}}]{2011A&A...535A..70P}
{Piffl}, T., {Williams}, M., \& {Steinmetz}, M. 2011, \aap, 535, A70, \dodoi{10.1051/0004-6361/201117474}

\bibitem[{{Portegies Zwart}(2000)}]{2000ApJ...544..437P}
{Portegies Zwart}, S.~F. 2000, \apj, 544, 437, \dodoi{10.1086/317190}

\bibitem[{{Quispe-Huaynasi} {et~al.}(2022){Quispe-Huaynasi}, {Roig}, {McDonald}, {Loaiza-Tacuri}, {Majewski}, {Wanderley}, {Cunha}, {Pereira}, {Hasselquist}, \& {Daflon}}]{2022AJ....164..187Q}
{Quispe-Huaynasi}, F., {Roig}, F., {McDonald}, D.~J., {et~al.} 2022, \aj, 164, 187, \dodoi{10.3847/1538-3881/ac90bc}

\bibitem[{{Ram{\'\i}rez} {et~al.}(2000){Ram{\'\i}rez}, {Sellgren}, {Carr}, {Balachandran}, {Blum}, {Terndrup}, \& {Steed}}]{2000ApJ...537..205R}
{Ram{\'\i}rez}, S.~V., {Sellgren}, K., {Carr}, J.~S., {et~al.} 2000, \apj, 537, 205, \dodoi{10.1086/309022}

\bibitem[{{Rau} \& {Pan}(2022)}]{2022ApJ...933...38R}
{Rau}, S.-J., \& {Pan}, K.-C. 2022, \apj, 933, 38, \dodoi{10.3847/1538-4357/ac7153}

\bibitem[{{Schlafly} {et~al.}(2019){Schlafly}, {Meisner}, \& {Green}}]{2019ApJS..240...30S}
{Schlafly}, E.~F., {Meisner}, A.~M., \& {Green}, G.~M. 2019, \apjs, 240, 30, \dodoi{10.3847/1538-4365/aafbea}

\bibitem[{{Schneider} {et~al.}(2020){Schneider}, {Burgasser}, {Gerasimov}, {Marocco}, {Gagn{\'e}}, {Goodman}, {Beaulieu}, {Pendrill}, {Rothermich}, {Sainio}, {Kuchner}, {Caselden}, {Meisner}, {Faherty}, {Mamajek}, {Hsu}, {Greco}, {Cushing}, {Kirkpatrick}, {Bardalez-Gagliuffi}, {Logsdon}, {Allers}, {Debes}, \& {Backyard Worlds: Planet 9 Collaboration}}]{2020ApJ...898...77S}
{Schneider}, A.~C., {Burgasser}, A.~J., {Gerasimov}, R., {et~al.} 2020, \apj, 898, 77, \dodoi{10.3847/1538-4357/ab9a40}

\bibitem[{{Scholz}(2024)}]{2024arXiv240210714S}
{Scholz}, R.-D. 2024, arXiv e-prints, arXiv:2402.10714, \dodoi{10.48550/arXiv.2402.10714}

\bibitem[{{Sch{\"o}nrich} {et~al.}(2010){Sch{\"o}nrich}, {Binney}, \& {Dehnen}}]{2010MNRAS.403.1829S}
{Sch{\"o}nrich}, R., {Binney}, J., \& {Dehnen}, W. 2010, \mnras, 403, 1829, \dodoi{10.1111/j.1365-2966.2010.16253.x}

\bibitem[{{Schultheis} {et~al.}(2015){Schultheis}, {Cunha}, {Zasowski}, {Garc{\'\i}a P{\'e}rez}, {Sellgren}, {Smith}, {Garc{\'\i}a-Hern{\'a}ndez}, {Zamora}, {Fritz}, {Anders}, {Allende Prieto}, {Bizyaev}, {Kinemuchi}, {Pan}, {Malanushenko}, {Malanushenko}, \& {Shetrone}}]{2015A&A...584A..45S}
{Schultheis}, M., {Cunha}, K., {Zasowski}, G., {et~al.} 2015, \aap, 584, A45, \dodoi{10.1051/0004-6361/201527027}

\bibitem[{{Schultheis} {et~al.}(2020){Schultheis}, {Rojas-Arriagada}, {Cunha}, {Zoccali}, {Chiappini}, {Zasowski}, {Queiroz}, {Minniti}, {Fritz}, {Garc{\'\i}a-Hern{\'a}ndez}, {Nitschelm}, {Zamora}, {Hasselquist}, {Fern{\'a}ndez-Trincado}, \& {Munoz}}]{2020A&A...642A..81S}
{Schultheis}, M., {Rojas-Arriagada}, A., {Cunha}, K., {et~al.} 2020, \aap, 642, A81, \dodoi{10.1051/0004-6361/202038327}

\bibitem[{{Sellwood} \& {Binney}(2002)}]{2002MNRAS.336..785S}
{Sellwood}, J.~A., \& {Binney}, J.~J. 2002, \mnras, 336, 785, \dodoi{10.1046/j.1365-8711.2002.05806.x}

\bibitem[{{Shen} {et~al.}(2018){Shen}, {Boubert}, {G{\"a}nsicke}, {Jha}, {Andrews}, {Chomiuk}, {Foley}, {Fraser}, {Gromadzki}, {Guillochon}, {Kotze}, {Maguire}, {Siebert}, {Smith}, {Strader}, {Badenes}, {Kerzendorf}, {Koester}, {Kromer}, {Miles}, {Pakmor}, {Schwab}, {Toloza}, {Toonen}, {Townsley}, \& {Williams}}]{2018ApJ...865...15S}
{Shen}, K.~J., {Boubert}, D., {G{\"a}nsicke}, B.~T., {et~al.} 2018, \apj, 865, 15, \dodoi{10.3847/1538-4357/aad55b}

\bibitem[{{Steinmetz} {et~al.}(2006){Steinmetz}, {Zwitter}, {Siebert}, {Watson}, {Freeman}, {Munari}, {Campbell}, {Williams}, {Seabroke}, {Wyse}, {Parker}, {Bienaym{\'e}}, {Roeser}, {Gibson}, {Gilmore}, {Grebel}, {Helmi}, {Navarro}, {Burton}, {Cass}, {Dawe}, {Fiegert}, {Hartley}, {Russell}, {Saunders}, {Enke}, {Bailin}, {Binney}, {Bland-Hawthorn}, {Boeche}, {Dehnen}, {Eisenstein}, {Evans}, {Fiorucci}, {Fulbright}, {Gerhard}, {Jauregi}, {Kelz}, {Mijovi{\'c}}, {Minchev}, {Parmentier}, {Pe{\~n}arrubia}, {Quillen}, {Read}, {Ruchti}, {Scholz}, {Siviero}, {Smith}, {Sordo}, {Veltz}, {Vidrih}, {von Berlepsch}, {Boyle}, \& {Schilbach}}]{2006AJ....132.1645S}
{Steinmetz}, M., {Zwitter}, T., {Siebert}, A., {et~al.} 2006, \aj, 132, 1645, \dodoi{10.1086/506564}

\bibitem[{{Sylos Labini} {et~al.}(2023){Sylos Labini}, {Chrob{\'a}kov{\'a}}, {Capuzzo-Dolcetta}, \& {L{\'o}pez-Corredoira}}]{2023ApJ...945....3S}
{Sylos Labini}, F., {Chrob{\'a}kov{\'a}}, {\v{Z}}., {Capuzzo-Dolcetta}, R., \& {L{\'o}pez-Corredoira}, M. 2023, \apj, 945, 3, \dodoi{10.3847/1538-4357/acb92c}

\bibitem[{{Vacca} {et~al.}(2003){Vacca}, {Cushing}, \& {Rayner}}]{2003PASP..115..389V}
{Vacca}, W.~D., {Cushing}, M.~C., \& {Rayner}, J.~T. 2003, \pasp, 115, 389, \dodoi{10.1086/346193}

\bibitem[{{van der Walt} {et~al.}(2011){van der Walt}, {Colbert}, \& {Varoquaux}}]{2011CSE....13b..22V}
{van der Walt}, S., {Colbert}, S.~C., \& {Varoquaux}, G. 2011, Computing in Science and Engineering, 13, 22, \dodoi{10.1109/MCSE.2011.37}

\bibitem[{{Vasiliev}(2019)}]{2019MNRAS.484.2832V}
{Vasiliev}, E. 2019, \mnras, 484, 2832, \dodoi{10.1093/mnras/stz171}

\bibitem[{{Virtanen} {et~al.}(2020){Virtanen}, {Gommers}, {Oliphant}, {Haberland}, {Reddy}, {Cournapeau}, {Burovski}, {Peterson}, {Weckesser}, {Bright}, {van der Walt}, {Brett}, {Wilson}, {Millman}, {Mayorov}, {Nelson}, {Jones}, {Kern}, {Larson}, {Carey}, {Polat}, {Feng}, {Moore}, {VanderPlas}, {Laxalde}, {Perktold}, {Cimrman}, {Henriksen}, {Quintero}, {Harris}, {Archibald}, {Ribeiro}, {Pedregosa}, {van Mulbregt}, \& {SciPy 1. 0 Contributors}}]{2020NatMe..17..261V}
{Virtanen}, P., {Gommers}, R., {Oliphant}, T.~E., {et~al.} 2020, Nature Methods, 17, 261, \dodoi{10.1038/s41592-019-0686-2}

\bibitem[{{Wenger} {et~al.}(2000){Wenger}, {Ochsenbein}, {Egret}, {Dubois}, {Bonnarel}, {Borde}, {Genova}, {Jasniewicz}, {Lalo{\"e}}, {Lesteven}, \& {Monier}}]{2000AAS..143....9W}
{Wenger}, M., {Ochsenbein}, F., {Egret}, D., {et~al.} 2000, \aaps, 143, 9, \dodoi{10.1051/aas:2000332}

\bibitem[{{W}es {M}c{K}inney(2010)}]{mckinney-proc-scipy-2010}
{W}es {M}c{K}inney. 2010, in {P}roceedings of the 9th {P}ython in {S}cience {C}onference, ed. {S}t\'efan van~der {W}alt \& {J}arrod {M}illman, 56 -- 61, \dodoi{10.25080/Majora-92bf1922-00a}

\bibitem[{{Williams} {et~al.}(2017){Williams}, {Belokurov}, {Casey}, \& {Evans}}]{2017MNRAS.468.2359W}
{Williams}, A.~A., {Belokurov}, V., {Casey}, A.~R., \& {Evans}, N.~W. 2017, \mnras, 468, 2359, \dodoi{10.1093/mnras/stx508}

\bibitem[{{Wilson} {et~al.}(2004){Wilson}, {Henderson}, {Herter}, {Matthews}, {Skrutskie}, {Adams}, {Moon}, {Smith}, {Gautier}, {Ressler}, {Soifer}, {Lin}, {Howard}, {LaMarr}, {Stolberg}, \& {Zink}}]{2004SPIE.5492.1295W}
{Wilson}, J.~C., {Henderson}, C.~P., {Herter}, T.~L., {et~al.} 2004, in Society of Photo-Optical Instrumentation Engineers (SPIE) Conference Series, Vol. 5492, Ground-based Instrumentation for Astronomy, ed. A.~F.~M. {Moorwood} \& M.~{Iye}, 1295--1305, \dodoi{10.1117/12.550925}

\bibitem[{{Wright} {et~al.}(2010){Wright}, {Eisenhardt}, {Mainzer}, {Ressler}, {Cutri}, {Jarrett}, {Kirkpatrick}, {Padgett}, {McMillan}, {Skrutskie}, {Stanford}, {Cohen}, {Walker}, {Mather}, {Leisawitz}, {Gautier}, {McLean}, {Benford}, {Lonsdale}, {Blain}, {Mendez}, {Irace}, {Duval}, {Liu}, {Royer}, {Heinrichsen}, {Howard}, {Shannon}, {Kendall}, {Walsh}, {Larsen}, {Cardon}, {Schick}, {Schwalm}, {Abid}, {Fabinsky}, {Naes}, \& {Tsai}}]{2010AJ....140.1868W}
{Wright}, E.~L., {Eisenhardt}, P.~R.~M., {Mainzer}, A.~K., {et~al.} 2010, \aj, 140, 1868, \dodoi{10.1088/0004-6256/140/6/1868}

\bibitem[{{Yu} \& {Tremaine}(2003)}]{2003ApJ...599.1129Y}
{Yu}, Q., \& {Tremaine}, S. 2003, \apj, 599, 1129, \dodoi{10.1086/379546}

\bibitem[{{Zhang} {et~al.}(2019){Zhang}, {Burgasser}, {G{\'a}lvez-Ortiz}, {Lodieu}, {Zapatero Osorio}, {Pinfield}, \& {Allard}}]{2019MNRAS.486.1260Z}
{Zhang}, Z.~H., {Burgasser}, A.~J., {G{\'a}lvez-Ortiz}, M.~C., {et~al.} 2019, \mnras, 486, 1260, \dodoi{10.1093/mnras/stz777}

\bibitem[{{Zhang} {et~al.}(2017){Zhang}, {Pinfield}, {G{\'a}lvez-Ortiz}, {Burningham}, {Lodieu}, {Marocco}, {Burgasser}, {Day-Jones}, {Allard}, {Jones}, {Homeier}, {Gomes}, \& {Smart}}]{2017MNRAS.464.3040Z}
{Zhang}, Z.~H., {Pinfield}, D.~J., {G{\'a}lvez-Ortiz}, M.~C., {et~al.} 2017, \mnras, 464, 3040, \dodoi{10.1093/mnras/stw2438}

\bibitem[{{Zhang} {et~al.}(2018){Zhang}, {Galvez-Ortiz}, {Pinfield}, {Burgasser}, {Lodieu}, {Jones}, {Mart{\'\i}n}, {Burningham}, {Homeier}, {Allard}, {Zapatero Osorio}, {Smith}, {Smart}, {L{\'o}pez Mart{\'\i}}, {Marocco}, \& {Rebolo}}]{2018MNRAS.480.5447Z}
{Zhang}, Z.~H., {Galvez-Ortiz}, M.~C., {Pinfield}, D.~J., {et~al.} 2018, \mnras, 480, 5447, \dodoi{10.1093/mnras/sty2054}

\end{thebibliography}
\bibliographystyle{aasjournal}

\end{document}